\documentclass[10pt,a4paper]{article}
\usepackage{amsmath,amssymb,latexsym}

\def\AFOUR{%
\setlength{\textheight}{8.5in}%
\setlength{\textwidth}{5.75in}%
\setlength{\topmargin}{-0.375in}%
\hoffset=-.5in%
\renewcommand{\baselinestretch}{1.17}%
\setlength{\parskip}{6pt plus 2pt}%
}

\AFOUR                                           

\expandafter\ifx\csname amssym.def\endcsname\relax \else\endinput\fi
\expandafter\edef\csname amssym.def\endcsname{%
       \catcode`\noexpand\@=\the\catcode`\@\space}
\catcode`\@=11
\def\undefine#1{\let#1\undefined}
\def\newsymbol#1#2#3#4#5{\let\next@\relax
 \ifnum#2=\@ne\let\next@\msafam@\else
 \ifnum#2=\tw@\let\next@\msbfam@\fi\fi
 \mathchardef#1="#3\next@#4#5}
\def\mathhexbox@#1#2#3{\relax
 \ifmmode\mathpalette{}{\m@th\mathchar"#1#2#3}%
 \else\leavevmode\hbox{$\m@th\mathchar"#1#2#3$}\fi}
\def\hexnumber@#1{\ifcase#1 0\or 1\or 2\or 3\or 4\or 5\or 6\or 7\or 8\or
 9\or A\or B\or C\or D\or E\or F\fi}

\font\tenmsa=msam10
\font\sevenmsa=msam7
\font\fivemsa=msam5
\newfam\msafam
\textfont\msafam=\tenmsa
\scriptfont\msafam=\sevenmsa
\scriptscriptfont\msafam=\fivemsa
\edef\msafam@{\hexnumber@\msafam}
\mathchardef\dabar@"0\msafam@39
\def\dashrightarrow{\mathrel{\dabar@\dabar@\mathchar"0\msafam@4B}}
\def\dashleftarrow{\mathrel{\mathchar"0\msafam@4C\dabar@\dabar@}}

\def\ulcorner{\delimiter"4\msafam@70\msafam@70 }
\def\urcorner{\delimiter"5\msafam@71\msafam@71 }
\def\llcorner{\delimiter"4\msafam@78\msafam@78 }
\def\lrcorner{\delimiter"5\msafam@79\msafam@79 }
\def\yen{{\mathhexbox@\msafam@55}}
\def\checkmark{{\mathhexbox@\msafam@58}}
\def\circledR{{\mathhexbox@\msafam@72}}
\def\maltese{{\mathhexbox@\msafam@7A}}
\def\circledS{{\mathhexbox@\msafam@73}}

\font\tenmsb=msbm10
\font\sevenmsb=msbm7
\font\fivemsb=msbm5
\newfam\msbfam
\textfont\msbfam=\tenmsb
\scriptfont\msbfam=\sevenmsb
\scriptscriptfont\msbfam=\fivemsb
\edef\msbfam@{\hexnumber@\msbfam}
\def\Bbb#1{{\fam\msbfam\relax#1}}
\def\widehat#1{\setbox\z@\hbox{$\m@th#1$}%
 \ifdim\wd\z@>\tw@ em\mathaccent"0\msbfam@5B{#1}%
 \else\mathaccent"0362{#1}\fi}
\def\widetilde#1{\setbox\z@\hbox{$\m@th#1$}%
 \ifdim\wd\z@>\tw@ em\mathaccent"0\msbfam@5D{#1}%
 \else\mathaccent"0365{#1}\fi}
\font\teneufm=eufm10
\font\seveneufm=eufm7
\font\fiveeufm=eufm5
\newfam\eufmfam
\textfont\eufmfam=\teneufm
\scriptfont\eufmfam=\seveneufm
\scriptscriptfont\eufmfam=\fiveeufm
\def\frak#1{{\fam\eufmfam\relax#1}}

\csname amssym.def\endcsname

\makeatletter
\def\section{\@startsection {section}{1}{\z@}{-3.5ex plus -1ex minus
 -.2ex}{2.3ex plus .2ex}{\large\sc}}
\def\subsection{\@startsection{subsection}{2}{\z@}{-3.25ex plus -1ex minus
 -.2ex}{1.5ex plus .2ex}{\normalsize\sc}}
\makeatother

\makeatletter
\@addtoreset{equation}{section}

\makeatother

\newcommand{\nc}{\newcommand}
\newcommand{\rnc}{\renewcommand}

\nc{\chap}[1]{{\clearpage}%
\begin{center}%
{\noindent\underline{\large\sc #1}}{\addcontentsline{toc}{section}{#1}}%
\end{center}%
{\vspace*{0.3cm}}}

\nc{\subs}[1]{{\vspace*{0.2cm}}%
{\noindent\underline{\small\sc
#1}}{\addcontentsline{toc}{subsubsection}{#1}}%
{\vspace*{0.2cm}}}

\nc{\be}{\begin{equation}}
\nc{\ee}{\end{equation}}
\nc{\bea}{\begin{eqnarray}}
\nc{\eea}{\end{eqnarray}}

\nc{\trac}[2]{{\textstyle\frac{#1}{#2}}}

\nc{\ex}[1]{\mbox{e}^{\,\textstyle#1}}

\nc{\CC}{\Bbb{C}}
\nc{\HH}{\Bbb{H}}
\nc{\PP}{\Bbb{P}}
\nc{\RR}{\Bbb{R}}
\nc{\ZZ}{\Bbb{Z}}
\nc{\II}{\Bbb{I}}
\nc{\EE}{\Bbb{E}}
\nc{\TT}{\Bbb{T}}
\nc{\DD}{\mathrm{I}\!\mathrm{D}}

\rnc{\a}{\alpha}
\rnc{\b}{\beta}
\rnc{\d}{\delta}
\nc{\ga}{\gamma}
\nc{\la}{\lambda}
\nc{\f}{\phi}
\nc{\e}{\eta}
\rnc{\c}{\chi}
\nc{\eps}{\epsilon}
\nc{\om}{\omega}
\nc{\Om}{\Omega}

\nc{\symx}{\circledS}
\newsymbol\smallsmile 1360
\newsymbol\smallfrown 1361
\nc{\ad}{\mathop{\mbox{ad}}\nolimits}
\nc{\tr}{\mathop{\mbox{tr}}\nolimits}
\nc{\Tr}{\mathop{\mbox{Tr}}\nolimits}
\nc{\Det}{\mathop{\mbox{Det}}\nolimits}
\rnc{\det}{\mathop{\mbox{det}}\nolimits}
\nc{\rk}{\mathop{\mbox{rk}}\nolimits}
\nc{\del}{\partial}
\nc{\diag}{\mathop{\mbox{diag}}\nolimits}
\nc{\ra}{\rightarrow}
\nc{\Ra}{\Rightarrow}
\nc{\LRa}{\Leftrightarrow}
\nc{\lra}{\leftrightarrow}
\nc{\ot}{\otimes}
\rnc{\ss}{\subset}
\nc{\nul}{\noindent\underline}
\nc{\non}{\nonumber\\}
\nc{\mat}[4]{\left(\begin{array}{cc}#1&#2\\#3&#4\end{array}\right)}
\rnc{\lg}{\frak{g}}
\nc{\G}[3]{\Gamma^{#1}_{\;{#2}{#3}}}
\nc{\nam}{\nabla_{\mu}}
\nc{\nan}{\nabla_{\nu}}
\nc{\dx}{\dot{x}}
\nc{\dxl}{\dot{x}^{\la}}
\nc{\dxm}{\dot{x}^{\mu}}
\nc{\dxn}{\dot{x}^{\nu}}
\nc{\ddx}{\ddot{x}}
\nc{\ddxm}{\ddot{x}^{\mu}}
\nc{\ddxn}{\ddot{x}^{\nu}}
\nc{\dxi}{\dot{\xi}}
\nc{\ddxi}{\ddot{\xi}}
\nc{\lsf}{\ell_s^{\mathrm{eff}}}
\nc{\lpf}{\ell_p^{\mathrm{eff}}}
\nc{\sqg}{\sqrt{g^{11}}}

\begin{document}


\vspace*{2cm}
\begin{center}
{\Large\sc DLCQ and Plane Wave Matrix Big Bang Models}
\end{center}
\vspace{0.2cm}

\begin{center}
{\large\sc Matthias Blau${}^a$} \textsc{and} 
{\large\sc Martin O'Loughlin${}^b$}\\[.8cm]
{\it ${}^a$ Institut de Physique, Universit\'e de Neuch\^atel, 
Breguet 1, Neuch\^atel, Switzerland}\\[.3cm]
{\it ${}^b$ University of Nova Gorica, Vipavska 13, 5000 Nova Gorica, Slovenia}
\end{center}

\vspace{1cm}

We study the generalisations of the Craps-Sethi-Verlinde matrix
big bang model to curved, in particular plane wave, space-times, beginning
with a careful discussion of the DLCQ procedure. Singular homogeneous
plane waves are ideal toy-models of realistic space-time singularities
since they have been shown to arise universally as their Penrose limits,
and we emphasise the role played by the symmetries of these plane waves
in implementing the
flat space Seiberg-Sen DLCQ prescription for these curved backgrounds.
We then analyse various aspects of the resulting matrix string Yang-Mills
theories, such as the relation between strong coupling space-time
singularities and world-sheet tachyonic mass terms. In order to have
concrete examples at hand, in an appendix we determine and analyse the
IIA singular homogeneous plane wave - null dilaton backgrounds.

\newpage

\vspace*{2.5cm}
\begin{small}
\tableofcontents
\end{small}

\newpage

\section{Introduction}

One of the main aims of string theory, as a theory of quantum gravity,
is to elucidate the nature and fate of space-time singularities. String
propagation in static space-times, such as time-independent orbifold
singularity backgrounds, is reasonably well understood. However,
comparatively little is known about string theory in non-trivial
time-dependent (and possibly singular) space-time backgrounds, the
time-dependence giving rise to rather basic problems in the very
formulation of string theory in such backgrounds, and the singularities
making questionable the validity or reliability of a perturbative 
approach to the problem. It is thus natural to appeal to more modern
non-perturbative formulations of string theory to gain some insight 
into these issues. An excellent summary of recent research along these 
lines, e.g.\ via the AdS/CFT correspondence or tachyon condensation,
can be found in \cite{berkooz}. 

One can also try to use non-perturbative matrix theory formulations
of M-theory \cite{bfss,susskind} or string theory \cite{motletc,dvv}
to address the fate of singularities. What had hampered progress along
these lines is the fact that these theories are quite complicated in
general (even weakly) curved backgrounds \cite{vrwt,schiappa}.  However,
it has recently been pointed out by Craps, Sethi and Verlinde (CSV)
\cite{csv} that one can find an explicit matrix string description of a
particular time-dependent IIA background, given in the string frame by
a flat metric with a linear null dilaton. This leads to a metric with
a null singularity either in the Einstein frame or upon lifting this
configuration to M-theory. The central observation of \cite{csv} is that
the dual matrix string gauge theory description of string theory in this
background is well-defined and weakly coupled close to the singularity. In
this regime the non-Abelian nature of the matrix-string coordinates
cannot be neglected and one thus tentatively arrives at a picture where
space-time geometry becomes non-commutative near a singularity, while
the emergence of a classical space-time at large distances from the
singularity has been confirmed by a 1-loop calculation \cite{csv2,craps}.
Subsequently, the CSV model has been extended and generalised in various
ways, e.g.\ to null brane backgrounds \cite{null1,null2,brunch} and 
certain plane wave metrics \cite{miaoli,chen,dm,ohta2}.

Here we will carefully revisit the extension of the CSV model to singular
plane wave backgrounds, not only because such geometries exhibit a perfect
combination of simplicity in construction together with interesting non-trivial
features such as time-dependence and a singularity, but also because 
the extension of the Seiberg-Sen DLCQ procedure
\cite{seiberg,sen,sen2} to curved backgrounds requires some care, and we feel
that previous treatements of this issue have not always been wholly satisfactory.

Moreover, as we will now argue, in the context of the matrix string
theory description of space-time singularities there is a privileged
class of IIA plane wave backgrounds, the homogeneous or scale-invariant
singular plane waves \cite{prt,mmhom} with a null dilaton.
Thanks to the special properties of these metrics they provide us with
a natural and interesting generalisation of the CSV background that also
permits an almost literal implementation of the generally accepted flat
space Seiberg-Sen procedure in these curved backgrounds.

We begin with the simple observation that the CSV 
IIA background \eqref{csv}, namely Minkowski space with a linear dilaton,
lifts to the 11-dimensional metric
\be
\label{csvm}
ds_{11}^2 = -2dudv + u\d_{ij}dy^idy^j + u^{-2} (dy)^2\;\;.
\ee
This exhibits the CSV background, first of all, as a
special case of an 11-dimensional plane wave, whose general form in Rosen
coordinates is $ds_{11}^2 = -2dudv + G_{\mu\nu}(u)dy^\mu dy^\nu$.
Secondly, and more specifically, 
the CSV background falls into the special class of plane wave metrics with 
the power-law behaviour
\be
\label{rcpl}
ds_{11}^2 = -2dudv + \sum_i u^{2n_i}(dy^i)^2 + u^{2b}(dy)^2\;\;.
\ee
It is precisely this class of metrics that was argued to provide an ideal
class of models of realistic space-time singularities, because these
singular scale-invariant homogeneous plane waves \cite{mmhom} were shown
in \cite{bbop1,bbop2,gpsi} to arise generically as the Penrose limits of
space-time singularities of power-law type \cite{SI}.\footnote{This is a
rather mild condition which essentially says that the singularity should
not be non-analytic in some suitably chosen coordinates.} Thus these
metrics are not just toy-models but true approximations (in the sense of
the Penrose-Fermi expansion \cite{bfw2,mbsw}) of space-time singularities
and, as such, provide a physically well-motivated generalisation of the
CSV background.

We thus need to understand how to extend the DLCQ construction
\cite{seiberg,sen} of matrix string theory, briefly reviewed in section
2.1 following \cite{sen2}, to these curved backgrounds. To
that end, we first carefully rephrase the procedure adopted by CSV
in the Seiberg-Sen framework (section 2.2). 
Having clarified what are precisely the
steps involved, in particular a null rotation (aligning the almost null
with a spacelike circle), a boost of the energies, and a subsequent
overall rescaling of the length scales of the theory, we then show that,
remarkably, the singular homogeneous plane waves are precisely such that
these operations can be implemented via isometries and homotheties of
the metric (section 2.3). 
We also resolve some ambiguities regarding the order in which
these isometries, scalings and the duality transformations implementing
the 9/11 flip \cite{dvv} of matrix string theory are to be performed.
This then allows us, as in \cite{csv}, to deduce the matrix string theory 
action from the expansion of the D-string DBI action (section 2.4)
and for its bosonic
part we obtain, schematically (for the precise expression see \eqref{src1}),
\be
\label{isrc1}
S\sim\int d^2\sigma \Tr \left(
\ex{2\phi(\tau)} F_{\alpha\beta}F^{\alpha\beta}
+ g_{ij}(\tau)D_\alpha X^i D^\alpha X^j
+
\ex{-2\phi(\tau)} g_{ik}(\tau)g_{jl}(\tau)[X^i,X^j] [X^k,X^l] \right)\;\;,
\ee
where the $g_{ij}$ are the components of the IIA Rosen coordinate plane
wave metric, and $\phi$ is the null dilaton. These actions, which are
the natural plane wave counterparts of ordinary Yang-Mills theory,
have appeared before in this context, e.g.\ in \cite{miaoli,chen,ohta2},
but both in our derivation of this DLCQ matrix string action and in the
analysis of the decoupling conditions (justifying the truncation to this
world-sheet Yang-Mills-like theory) our treatment differs significantly
from that of previous publications.

Having obtained this action, we then analyse some of its properties.
We comment on the fermionic terms in the action in section 3.1.  In
section 3.2 we show that the space-time coordinate transformation from 
Rosen to
Brinkmann coordinates results in a rather non-obvious equivalence between
two apparently very different matrix string Yang-Mills
theories (namely the above action, with time-dependent couplings for
the scalar fields on the one hand, and the Brinkmann action \eqref{sbc}
with time-independent couplings but time-dependent mass terms instead).
This illustrates that the above action correctly captures the target
space-time geometry. The brief discussion of this equivalence presented
here will be supplemented by a more detailed discussion in \cite{blrcbc},
where we also extend it to 3-algebra actions such as those that appear in
the BLG multiple M2-brane actions \cite{bl,gu,lor3},

In section 3.3 we investigate the possibility to absorb the time-dependent
couplings into the world-sheet metric, as in \cite{csv}, finding
one special case where one ends up with a (1+1) de Sitter world-sheet,
and in section 3.4 we highlight the usefulness of the (more space-time
covariant) Brinkmann representation of the action by showing how the 
space-time nature of the singularity (whether it is strongly or weakly
coupled) is reflected in the mass terms of the world-sheet theory (strong coupling 
requiring at least one `tachyonic' scalar). 

We briefly summarise some useful facts about the geometry of plane waves
in appendix A, and in appendix B we classify and analyse the plane wave -
null dilaton backgrounds that arise upon reduction from the 11d plane wave
power-law metrics \eqref{rcpl}. 

One subject we do not address here is that of classical solutions in these
models, such as fuzzy spheres, and the possible role that they may play
in understanding the evolution away from/towards the null singularity.
These have been studied before in this context, e.g.\ in \cite{dm,chen},
and some interesting new work in this direction will appear in \cite{dfsw}. 

\section{DLCQ and Singular Homogeneous Plane Wave Backgrounds}

\subsection{Quick Recap of the Seiberg-Sen Argument}

In the Seiberg-Sen argument \cite{seiberg,sen}
(as well as in its CSV variant \cite{csv,craps} we will
discuss below) a central role is played by the transformation which
relates the compactification on a (vanishingly) small spacelike circle
with radius $R_s\ra 0$ 
to the DLCQ compactification on a null circle
with fixed radius $R$. Concretely, one considers the metric
\be
ds^2 = - 2dy^+dy^- + \ldots = -(dy^0)^2 + (dy^9)^2 + \ldots 
\ee
where $y^\pm = (y^0\pm y^9)/\sqrt{2}$. 
In order to realise the lightlike identification
$y^- \sim y^- \pm 2\pi R$ as a limit of 
standard spacelike compactifications, 
one considers the boost
\be
\label{xppm}
x^{\pm} \equiv y^{\prime\pm} = \ex{\pm\beta}y^\pm
\ee
and requires that in the boosted/primed coordinates the identification is
\be
\label{sli}
x^{9} \sim x^{9} + 2\pi R_s\qquad\qquad x^0\sim x^0\;\;.
\ee
A convenient choice is
$\ex{\beta} = \sqrt{2} \frac{R}{R_s}$,
corresponding to the simple identification 
\be
\label{ypp}
y^+ \sim y^+ + \pi R_s^2/R\qquad\qquad y^- \sim y^- - 2\pi R
\ee
of the unboosted lightcone coordinates $y^\pm$ (with this choice of $\beta$
the periodicity of $y^-$ is fixed, i.e.\ independent of $R_s$). 
Since the lightcone coordinates transform under a boost as in \eqref{xppm},
the momenta transform as
$p^{\prime}_{\pm} = \ex{\mp \beta}p_{\pm}$.
In particular, a state with $N$ units of momentum  in the compact
$x^9$-direction, $p^{\prime}_9 = N/R_s$, leads in the limit $R_s\ra 0$
to a state with $N$ units of lightcone momentum 
\be
p^+ \equiv -p_- = \frac{N}{R}\;\;,
\ee
since
\be
\frac{N}{R_s} = p^\prime_9 = \trac{1}{\sqrt{2}}(p^\prime_+-p^\prime_-) 
= \trac{1}{\sqrt{2}}(\ex{-\beta} p_+-\ex{\beta}p_-) 
\stackrel{R_s\ra 0}{\longrightarrow} -\frac{R}{R_s}p_-
\ee
Now, following \cite{sen2}, 
one defines the DLCQ Hamiltonian $H_N^{DLCQ}(m,R)$ 
(in a sector with $N$ units of lightcone momentum 
and characterised by, say, a mass scale $m$) as the limit
\be
H_{N}^{DLCQ}(m,R) := \lim_{R_s\ra0} i\del_{y^+}\;\;,
\ee
with $m$ and $R$ fixed,
since in this limit the spacelike identification 
\eqref{sli} becomes the lightlike identification 
$y^-\sim y^- + 2\pi R$.
In terms of the boosted coordinates $x^{\mu} = y^{\prime\mu}$
one has
\be
\label{elc2}
E_{lc}\equiv i\del_{y^+}  =
\frac{1}{\sqrt{2}}\ex{\beta}(i\del_{x^0}+ i\del_{x^9}) 
= \frac{R}{R_s}(E^\prime - p^\prime_9)\;\;.
\ee
The term in brackets on the rhs is the total energy of the system minus the
background energy $p^\prime_9 = N/R_s$, Sen's \cite{sen,sen2} KK
Hamiltonian 
$H_N(m,R_s) := E^\prime - p^\prime_9$.
In the present context, it is also useful to think of
this linear combination in the limit $p^\prime_9 = N/R_s \ra\infty$ as the
non-relativistic infinite momentum frame Hamiltonian 
\be
\label{imf1}
E^\prime = \sqrt{(p^\prime_9)^2 + \vec{p}^{\,\prime 2} + m^2} \quad 
\Ra\quad  E^\prime - p_9^\prime = 
\frac{\vec{p}^{\,\prime 2} + m^2}{2p_9^\prime} +
\mathcal{O}((p_9^\prime)^{-3})
\;\;.
\ee
Thus one has 
\be
H_N^{DLCQ}(m,R) = \lim_{R_s\ra 0} \frac{R}{R_s}H_N(m,R_s)\;\;.
\ee
In order to identify the rhs, and eliminate the singular prefactor, 
one now makes the observation that, on purely
dimensional grounds, if one rescales all mass scales by a factor $\lambda$,
and all length scales by a factor $\lambda^{-1}$, then the Hamiltonian will 
also scale as $\lambda$. 
Thus, with $\lambda = R/R_s$ one has 
\be
\frac{R}{R_s}H_N(m,R_s) = H_N (\frac{R}{R_s}m,\frac{R_s^2}{R})\equiv
H_N(\hat{m},\hat{R}_s)\;\;,
\ee
and one has now concretely realised the DLCQ Hamiltonian 
as the limit
\be
H_N^{DLCQ}(m,R) = \lim_{\hat{R}_s\ra 0} H_N(\hat{m},\hat{R}_s)
\ee
of standard KK Hamiltonians of a family of new theories with
mass scale 
$\hat{m} = (R/R_s) m$
and spatial radius 
$\hat{R}_s = R_s^2/R$,
with $m$ and $R$ held fixed.\footnote{Note that, while in terms of $R_s$ 
one has $R_s \hat{m} = R m$
(this really is just a change of scale, the dimensionless quantity $Rm$ being
kept fixed), in terms of $\hat{R}_s$ one has the Seiberg relation
$ \hat{R}_s \hat{m}^2 = R m^2$
\cite{seiberg}.}

This much is completely general, and pure kinematics. One can now apply
this to $M$-theory with $m=m_p$ the Planck mass and
$R_s=R_{11}$. In terms of the scaled IIA parameters
$\hat{g}_s$ and $\hat{\ell}_s$ (string coupling and string length),
this is then precisely the DKPS \cite{dkps}
D0-brane weakly coupled ($\hat{g}_s\ra
0$) field theory ($\hat{\ell}_s \ra 0$) limit, in which the Hamiltonian
reduces to YM quantum mechanics with the finite YM coupling $g^2_{YM}
= \hat{g}_s\hat{\ell}_s^{-3}$, leading to the BFSS matrix theory \cite{bfss}.

One can equally well apply this prescription to IIA string
theory \cite{motletc,dvv,lifshytz}.
Thus, starting off with a IIA theory with string scale $m_s$
and null circle with radius $R$, this is described by the limit of a
$\widehat{IIA}$-theory with string scale $\hat{m}_s = (R/R_s) m_s$
and radius $\hat{R}_s = R_s^2/R$, and with the same (dimensionless)
coupling constant $\hat{g}_s = g_s$. To identify this theory, one can
lift it to 11-dimensions. Then one sees that this theory is described
by the above YM matrix theory with an additional transverse circle,
bettter described by (1+1)-dimensional YM theory on the dual cylinder
with constant YM coupling constant $g_{YM}^2 \sim (R/g_s\ell_s^2)^2$
and constant radius $R_{D1} = \ell_s^2/R$ on which the T-dual D1-branes
are wrapped. This is the matrix string DLCQ description of IIA string theory.

\subsection{Adapting the Seiberg-Sen Argument to the CSV Setting}

In \cite{csv,craps}, a variant of the Seiberg-Sen argument was introduced, 
in which the lightlike compactification is related to a limit of spacelike
compactifications along a direction transverse to the lightcone. The
motivation for this was the fact that in the CSV model one has a null linear
dilaton, which is not compatible with the periodic identification of $y^+$ in
\eqref{ypp}. In this section, in order to prepare the ground for the
generalisation to curved backgrounds, we will explain the precise relationship
between the procedure adopted by CSV and the Seiberg-Sen argument of the
previous section. 

The point of departure this time is a metric of the form
\be
ds^2 = - 2dy^+dy^- + (dy^1)^2 + \ldots,
\ee
where we seek a Lorentz tranformation such that an almost null identification 
in this background is mapped to the manifestly spatial identification
$\tilde{x}^{1} \sim \tilde{x}^{1} + 2\pi R_s$.
If one does (for the time being) not touch $y^+$, this leaves null
rotations which have the general form
\be
\label{xty}
y^- = \tilde{x}^{-} + \alpha \tilde{x}^1 + (\alpha^2/2) \tilde{x}^+
\qquad\qquad
y^1 = \tilde{x}^{1} + \alpha \tilde{x}^+\;\;.
\ee
The identification then becomes
\be
\label{ani}
y^1 \sim y^1 + 2\pi R_s\qquad\qquad y^-\sim y^- + 2\pi \alpha R_s
\ee
Thus the obvious choice for $\alpha$ (leading, as in section 2.1, to 
an $R_s$-independent periodicity of $y^-$), is
$\alpha = \pm R/R_s$. Once again, 
a state with $N$ units of momentum  in the compact
direction, $\tilde{p}_1 = \mp N/R_s$, 
is mapped in the limit $R_s\ra 0$
to a state with $N$ units of lightcone momentum 
$p^+ = N/R$, since
\be
\tilde{p}_1 = p_1 + \alpha p_-  = p_1 \pm \frac{R}{R_s}p_-
\stackrel{R_s\ra 0}{\longrightarrow}
\pm \frac{R}{R_s}p_-\;\;.
\ee
Therefore, we can again define the DLCQ Hamiltonian as the limit
\be
\label{dlcqdef}
H_{N}^{DLCQ}(m,R) := \lim_{R_s\ra0} i\del_{y^+}\;\;,
\ee
this time with the almost null identification \eqref{ani}, and the aim is now
to rewrite this as a well-defined limit of standard Hamiltonians. 
Here it is important to note that, while in the Seiberg-Sen procedure the
boosting of the energies, as in \eqref{elc2},
was an automatic consequence of aligning an almost lightlike
with a manifestly spacelike direction, here this is not the case. 
Rather, under the null rotation, which achieves this alignment all by itself,
the lightcone energy transforms as
\be
\label{etlc}
\tilde{E}_{lc} = E_{lc} - \alpha p_1 - (\alpha^2/2)p_-
\ee
and is thus not becoming small. Thus one can anticipate that
this boost must still be performed seperately in order
to arrive at the vanishing
energies that permit a decoupling limit argument. Moreover, the Seiberg-Sen
boost had the added bonus that it automatically led to the appropriate
background subtracted KK Hamiltonian 
$H_N =
E^\prime-p_9^\prime$. Using the CSV prescription, this
is not automatically the case, the background momentum arising from the
$x^1$-direction, which is not part of the lightcone directions. However, 
we will see that with a judicious choice of null rotation and boost
parameters we will once again be able to relate the DLCQ Hamiltonian to the
relevant KK Hamiltonian, namely $E^\prime - p^\prime_1$. 

To make this more explicit, we go to the
adapted null-rotated coordinate system
$\tilde{x}^{\mu}$  \eqref{xty}, and 
also perform a further boost isometry, with
parameter $\gamma > 0$, to the coordinates $y^{\prime \mu}=x^{\mu}$, with
\be
\label{csvboost2}
x^\pm = \gamma^{\pm 1} \tilde{x}^\pm\;\;.
\ee
Then one has the relation
\be
\label{elc3}
i\del_{y^+}= \gamma i \del_{x^+} - \alpha i \del_{x^1}+ \gamma^{-1}\alpha^2/2
\;i\del_{x^-}\quad \LRa\quad E_{lc} =
\gamma (E^{\prime}_{lc} + (\alpha/\gamma) p^\prime_1 -(\alpha^2/2\gamma^2)
p^\prime_-)
\ee
To unravel this, let us write the lightcone energies and momenta in terms of
the ordinary energy $E^\prime=i\del_{x^0}$ and $p_9^\prime=-i\del_{x^{9}}$,
\be
\label{elc10}
E_{lc} = \frac{\gamma}{\sqrt{2}}\left( (1+ \alpha^2/2\gamma^2) E^\prime -
(1-\alpha^2/2\gamma^2)p^\prime_9 + (\sqrt{2}\alpha/\gamma) p^\prime_1 \right)
\;\;.
\ee
We see that we can eliminate the annoying $p^\prime_9$ from this expression
by choosing $\alpha$ and $\gamma$ such that $\alpha^2/2\gamma^2=1$. 
This also
has the consequence that $p^\prime_1$ appears with the same coefficient as
$E^\prime$. 
Specifically, choosing 
\be
\label{ag1}
\alpha = -\frac{R}{R_s}\qquad\qquad \gamma = \frac{R}{\sqrt{2}R_s}\;\;,
\ee 
one finds the simple result 
\be
\label{elc8}
E_{lc} = \frac{\gamma}{\sqrt{2}}( 2 E^\prime -
2 p^\prime_1 ) = \frac{R}{R_s}(E^\prime - p^\prime_1)
\ee
which is the precise analogue of the expression 
\eqref{elc2}
that appears in the argument based on the Seiberg boost. The expression in
brackets is Sen's KK Hamiltonian, and thus one can now repeat
\textit{verbatim} the arguments of the previous section to deduce that
\be
H_N^{DLCQ}(m,R) = 
\lim_{R_s\ra 0} \frac{R}{R_s}H_N(m,R_s)
=\lim_{\hat{R}_s\ra 0} H_N(\hat{m},\hat{R}_s)\;\;.
\ee
The change of scale
involved in the last equality means that 
the relation \eqref{elc8} between the 
lightcone energy and the boosted energy becomes the statement that 
the original lightcone energy is now equal to the
background subtracted energy in the final (YM-like in the string context) 
rescaled theory,  
\be
\label{elc9}
E_{lc} = \hat{E} - \hat{p}_1\;\;.
\ee
It is convenient, not only for book-keeping purposes,
to concretely implement this change of scale
by the scaling
\be
\label{xgs}
\hat{x}^{\mu} = \frac{R_s}{R}x^{\mu} \qquad\qquad d\hat{s}^2 =
\left(\frac{R_s}{R}\right)^2 ds^2 \;\;.
\ee
of the coordinates and the metric. In particular, the new energies and
momenta are now related to the old ones by $\hat{E} = (R/R_s) E^{\prime}$
etc., leading to \eqref{elc9}.  Moreover, the concomitant rescaling of the
metric has precisely the effect that other length scales in the problem,
like transverse radii, are also automatically rescaled appropriately,
i.e.\ a circle of proper radius $\rho$ with respect to the metric $ds^2$
has proper radius $\hat{\rho} = (R_s/R) \rho$ with respect to the metric
$d\hat{s}^2$.

We should note here that, while the above choice of parameters \eqref{ag1}
leads to a nice cancellation among various terms, and thus to the simple
final result \eqref{elc9}, there is considerable leeway in the choice
of parameters if one is only interested in some gauge-fixed energy
fluctuations. In particular if, in the string context and as in \cite{csv},
one gauge fixes $x^+$ and $x^1$, 
this amounts to setting the fluctuations $\d \hat{p}_-$ and
$\d \hat{p}_1$ to zero. This implies, in particular, that
$\delta \hat{E}_{lc} \sim \d \hat{E}$,
so that at the level of the gauge-fixed fluctuations \eqref{elc10}
leads to 
\be
\label{elc11}
\d E_{lc} \sim \d \hat{E} \qquad\forall\; \gamma \sim \alpha = \pm
\frac{R}{R_s}
\ee
(with a finite proportionality factor),
which thus also gives a direct
relationship between the lightcone and YM (fluctuation) energies.  
In particular, the choice adopted by CSV is 
\be
\label{agcsv}
\alpha = \gamma =  \frac{R}{R_s}\;\;,
\ee 
leading to the (apparently less attractive) result
\be
E_{lc} = (\hat{E}_{lc} + \hat{p}_1 - \trac{1}{2}\hat{p}_-)
\;\;,
\ee
which, however, reduces to \eqref{elc11} at the level of fluctuations.
This is good enough.
Below, when
discussing the generalisation of this procedure to curved backgrounds,
we will similarly make use of this freedom (and adopt the CSV choice
\eqref{agcsv}) to define an appropriate fluctuation Hamiltonian since, 
in general, in any case
there will be no choice of constant parameters $\alpha,\gamma$
that gives on the nose the appropriate curved space 
analogue \eqref{imf2} of (\ref{imf1},\ref{elc9}).

\subsection{Extending the Seiberg-Sen-CSV DLCQ to Plane Wave Backgrounds: \\
the Privileged Role of Singular Homogeneous Plane Waves}

We will now discuss the generalisation of the CSV derivation of matrix
string theory to curved (in particular plane wave) backgrounds. To
that end, let us first take stock of the steps involved in the CSV
procedure.  As we saw above, in the case of a flat background (possibly
supplemented by a null dilaton \cite{csv}), the derivation can be
concretely implemented by the following three steps:

\begin{enumerate}
\item As a first step, one performs the purely kinematical operation of 
passing to adapted coordinates by a coordinate transformation
that aligns the almost null with a small space-like circle. In the case
of the flat metric this is accomplished by a null rotation Lorentz isometry.

\item Next one performs a dynamical transformation (acting non-trivially on the
lightcone time  coordinate $y^+$) which has the effect of rescaling the
energies.  In the case
of the flat metric this is accomplished by a boost Lorentz isometry.

\item Finally, this boost is accompanied by a rescaling of the coordinates
(and the metric), implementing the Seiberg-Sen change of mass/length scales.
In the case
of the flat metric this is accomplished by a uniform scaling of the Minkowski
coordinates, which is a homothety (constant conformal rescaling) of the
metric.
\end{enumerate}

We will now show that this prescription can be implemented almost
literally for a special class of plane wave metrics (in the above,
replace ``flat'' by ``singular homogeneous plane wave'', ``Minkowski''
by ``Brinkmann'', and eliminate the word ``Lorentz'').

To set the stage, we first consider a general plane wave metric, which in
Rosen coordinates takes the form
\be
\label{3pwrc}
ds^2 = -  2dy^+ dy^- + g_{ij}(y^+)dy^i dy^j\;\;.
\ee
Note that this metric has manifest translational isometries in the lightcone
$y^-$- and transverse $y^i$-directions, and we thus begin
with the almost null identification 
\be
\label{yid}
(y^+,y^-,y^1,y^m) \sim (y^+, y^-+2\pi R, y^1 + 2 \pi \epsilon R,y^m)\;\;.
\ee
where we have set
$R_s=\epsilon R$. As before, we would like to perform a coordinate 
transformation (and ideally an isometry) $y^\mu \ra \tilde{x}^{\mu}$, 
in terms of which the above identification simply reads 
\be
\label{xtid}
\tilde{x}^1 \sim \tilde{x}^1 + 2\pi \epsilon R\;\;.
\ee
Now in addition to the manifest transverse translational isometries, 
any plane wave metric \eqref{3pwrc} has dual hidden translational symmetries
generated by the transverse Killing vectors $P^{(i)}$ \eqref{api}. 
In particular, $P^{(1)}$
generates the transformation (null rotation)
\be
\label{boost3}
(y^+,y^-,y^1,y^m) = (\tilde{x}^+,\tilde{x}^- + \alpha\tilde{x}^1 +
\alpha^2 h^{11}(\tilde{x}^+)/2 ,
\tilde{x}^1 + \alpha h^{11}(\tilde{x}^+),
\tilde{x}^m + \alpha h^{1m}(\tilde{x}^+))\;\;,
\ee
(where $h^{ik}(y^+) = \int^{y^+}\!\!du\; g^{ik}(u)$),
which indeed accomplishes \eqref{xtid} for the choice $\alpha = \epsilon^{-1}$.

Thus step 1 can be implemented via isometries for any plane wave metric.
Step 2 requires an isometry involving transformations of $y^+$. Generic plane
waves do not possess any such isometries. However, there are precisely
two classes of plane waves with such an extra isometry \cite{mmhom}. In one,
this isometry involves a shift in $y^+$, in the other the isometry 
is realised not by a shift but by a scaling of $y^+$ (accompanied 
by some transformation of the other coordinates). In the present context, it is
evidently this latter class of \textit{singular homogeneous plane waves}
(SHPWs) that we are interested in. The prototypical example of such SHPWs
are plane wave metrics with a power-law behaviour in Rosen coordinates, 
\be
\label{3plrc}
ds^2 = - 2dy^+dy^- + \sum_i (y^+)^{2m_i} (dy^i)^2\;\;.
\ee
This isometry is more manifest in Brinkmann coordinates, in which the above
metric takes the form
\be
\label{3bcbn1}
ds^2 = -2dz^+dz^- + \sum_a m_a(m_a-1)(z^a)^2 \frac{(dz^+)^2}{(z^+)^2} 
+ \sum_a (dz^a)^2\;\;.
\ee
This metric clearly possesses 
the isometry is $(z^+,z^-)\ra (\lambda z^+,\lambda^{-1}z^-)$ and, translated
back to Rosen coordinates, 
this isometry is given by
$(y^+,y^-,y^i) \ra (\lambda y^+,\lambda^{-1}y^-,\lambda^{-m_i}y^i)$.
In particular, the Lorentz boost isometry \eqref{csvboost2}
can be generalised to the isometry
\be
\label{boost4}
(\tilde{x}^+,\tilde{x}^-,\tilde{x}^i) = (\gamma^{-1}
x^+,\gamma x^-,\gamma^{m_i}x^i)\;\;,
\ee
and we will eventually choose $\gamma = \alpha = \epsilon^{-1}$, in
accordance with \eqref{agcsv}.

Now let us consider Step 3, the scaling. First of all we 
note that any plane wave 
\be
ds^2 = -2dz^+dz^- + A_{ab}(z^+) z^a z^b (dz^+)^2 + (dz^a)^2
= -2dx^+dx^- + g_{ij}(x^+)dx^i dx^j
\ee
has the homothety (conformal isometry with a constant factor)
\be
\label{homo1}
(z^+,z^-,z^a) \ra (z^+,\lambda^2 z^-, \lambda z^a)\quad\LRa\quad
(x^+,x^-,x^i) \ra (x^+,\lambda^2 x^-, \lambda x^i)
\ee
under which $ds^2 \ra \lambda^2 ds^2$. Even though this scales the
metric, it does not uniformly rescale the energies / length scales of the
theory as e.g.\ $x^+$ does not scale. However, precisely when 
$A_{ab}(z^+) \sim (z^+)^{-2}$, as in \eqref{3bcbn1}, 
there is another homothety, 
namely the uniform rescaling of the coordinates
\be
(z^+,z^-,z^a) \ra \lambda (z^+,z^-, z^a)\;\;.
\ee
(this is a combination of the first homothety with the boost). 
Since Brinkmann coordinates are Fermi coordinates \cite{bfw2} and
thus a direct measure of geodesic distances,
this is indeed a physical scale transformation.
Thus in this
case, one can complete the Seiberg-CSV procedure in a natural way by a
rescaling of the coordinates and the metric exactly as in \eqref{xgs}, 
\be
\label{bcscaling}
\hat{z}^{\mu} = \epsilon z^\mu
\qquad\qquad d\hat{s}^2 = \epsilon^2 ds^2 \;\;.
\ee
In Rosen coordinates, this transformation of the coordinates reads
\be
\label{rcscaling}
(\hat{x}^+,\hat{x}^-,\hat{x}^i) = (\epsilon x^+,\epsilon x^-,
\epsilon^{1-m_i} x^i)\;\;.
\ee
Thus, to summarise, there appears to be a straightforward and very natural
extension of the Seiberg-CSV procedure to scale-invariant plane waves, in
which the flat space 
Lorentz transformations (null rotation, boost) are implemented
by isometries of the metric, and the Seiberg scaling is realised
by a uniform scaling of the Brinkmann coordinates 

We are thus now in the position to define,  
for a power-law plane wave metric of the form 
\be
ds^2 = -2dy^+ dy^- + g_{11}(y^+) (dy^1)^2 + \ldots
= -2dy^+ dy^- + (y^+)^{2m_1} (dy^1)^2 + \ldots
\ee
the DLCQ Hamiltonian, as before, via
\be
H_N^{DLCQ}(m,R) := \lim_{R_s\ra 0} i\del_{y^+}\;\;.
\ee
Performing the null
rotation isometry \eqref{boost3} with parameter $\alpha$, the boost isometry 
\eqref{boost4} with parameter $\gamma$, and the homothety \eqref{rcscaling}
with parameter $\epsilon$ (momentarily treating these parameters as
independent), one finds, as the generalisation of \eqref{elc3} (and with the
final scaling already performed) 
\be
\del_{y^+}= 
\epsilon\gamma\left[\del_{\hat{x}^+} - (\alpha/\gamma) (\epsilon\gamma)^{m_1}
g^{11}(\hat{x}^+)
\;\del_{\hat{x}^1}+ (\alpha^2/2\gamma^2) (\epsilon\gamma)^{2m_1}
g^{11}(\hat{x}^+)
\;\del_{\hat{x}^-}\right]
\ee
We see that this has a finite limit as $\epsilon \ra 0$ ($R_s\ra 0$),
provided that $\gamma \sim \alpha \sim \epsilon^{-1}$. We also see that, for
non-contant $g_{11}$ there is no choice of constant parameters $\alpha,
\gamma$ for which the rhs reduces directly to the 
background subtracted energy, namely $\hat{E} -
\sqrt{g^{11}(\hat{x}^+)}\hat{p}_1$, arising from the 
appropriate generalisation of \eqref{imf1},
\be
\label{imf2}
\hat{E} = \sqrt{g^{11}(\hat{x}^+)(\hat{p}_1)^2 + \vec{\hat{p}}^{2} + m^2} \quad 
\Ra\quad  \hat{E} - \sqrt{g^{11}(\hat{x}^+)}\hat{p}_1 = 
\frac{\vec{\hat{p}}^{2} + m^2}{2\sqrt{g^{11}(\hat{x}^+)}\hat{p}_1} + \ldots
\ee
However, as discussed at the end of section 2.2, it is still meaningful to
use the above equation at the level of gauge-fixed energy fluctuations for 
any choice of $\alpha,\gamma,\epsilon$, subject to the condition 
$\gamma \sim \alpha \sim \epsilon^{-1}$, and we will henceforth make the
simple choice \eqref{agcsv}
\be
\label{agfinal}
\alpha = \gamma = \epsilon^{-1} = \frac{R}{R_s}\;\;,
\ee
leading to $\d E_{lc} \sim \d \hat{E}$.
With this choice of parameters, the combined action of the boost
\eqref{boost4} and the rescaling \eqref{rcscaling} is the transformation
\be
\label{pensc}
(\tilde{x}^+,\tilde{x}^-,\tilde{x}^i) = (\hat{x}^+, \epsilon^{-2} \hat{x}^-,
\epsilon^{-1}\hat{x}^i)\;\;,
\ee
which, as noted before, is actually a homothety \eqref{homo1} for
\textit{any} plane wave metric. In particular, the original lightcone time
coordinate $y^+$ is equal to the ``Yang-Mills'' lightcone gauge time variable
$\hat{x}^+=\tau$. Moreover, combined with the rescaling
\eqref{bcscaling} of the metric, this is precisely the scaling that
defines the Penrose plane wave limit \cite{penrose,gueven,bfp}. In the present
case of plane waves, the Penrose limit leaves the metric invariant,
and this is precisely as it should be since the DLCQ procedure
should ideally not deform the metric (or other background fields).

Obviously, once one thinks of this combined transformation,
one is naturally led to explore the relationship between DLCQ and the
Penrose limit for more general backgrounds, e.g.\ as in \cite{shomer}
or \cite{sheikh,torabian}. However, it is not clear to us whether the
result should then really be thought of as the DLCQ of a theory in the
original background (this appears to be the point of view adopted e.g.\
in \cite{sheikh}) rather than as a DLCQ of a theory in some Penrose
plane wave limit (depending on a choice of null geodesic) of the original
background.

As a final remark, we should also point out that in some of the post-CSV
literature dealing with the extension of the CSV model to curved
backgrounds (typically plane waves, even though not always recognised as
such, written in almost-Rosen coordinates like
(\ref{cpwrc},\ref{cpwrc2})), the boost and scale transformation of the
Seiberg-Sen procedure were implemented by the naive flat space boost
\be
(y^+,y^-,y^i) \ra (\epsilon y^+,\epsilon^{-1} y^-,y^i) 
\ee
and the naive flat space scaling
\be
(y^+,y^-,y^i) \ra \epsilon^{-1} (y^+,y^-,y^i) \;\;.
\ee
Now the former is not an isometry of a plane wave metric (not even when it is
of the power-law, singular homogeneous, type), and the latter
is neither a homothety nor a physical scale transformation of the metric 
(Rosen coordinates are not a measure of proper physical distance in
space-time). Nevertheless, the combined action
of these two transformations happens to be identical to the combined action
\eqref{pensc} resulting from the boost isometry
\eqref{boost4} and the physical scale transformation \eqref{rcscaling}, 
and thus in the present case one can 
get away with this. However, it should be clear
from what we have said that conceptually at least this appears to be
an incorrect
implementation of DLCQ, or at least one that requires further justification.
Our more systematic
treatment of the DLCQ will also
lead to a quite different, and significantly simpler, analysis of decoupling
conditions and related issues.

\subsection{9/11 Flip, DBI Expansion, and Decoupling}

We have now prepared the ground for the DLCQ of IIA string theory
in singular homogeneous plane wave - null dilaton backgrounds such
as those determined in appendix B, \eqref{bn1}. We assume that we
have already performed the null rotation $y^{\mu}\ra\tilde{x}^{\mu}$ 
aligning the almost null circle with the spatial direction $\tilde{x}^1$,
such that
$\tilde{x}^1 \sim \tilde{x}^1 + 2\pi R_s$, and we focus on a sector
with $N$ units of momentum $\tilde{p}_1 = N/R_s$. 

Via the procedure outlined at the end of section 2.1 (boosting
and scaling, lifting to M-theory along $x^{11}$ with scaled radius
$\hat{R}_{11}=\epsilon \ell_s g_s$, $\ell_s$ and $g_s$ denoting the
original IIA string length and string coupling respectively, then
reducing along the scaled circle with radius $\hat{R}_1 = \hat{R}_s =
\epsilon^2 R$, and performing a T-duality along $x^{11}$), one arrives at
a definition of the DLCQ of IIA string theory in terms of a (decoupling)
limit of IIB string theory in a sector with $N$ D1-branes.

Alternatively, one can use the 9/11 flip \cite{motletc,dvv} (more
appropriately called a 1/11 flip in the present context) to arrive at
the same theory (in entirely 10-dimensional terms) by performing first
the boost and scaling, then a T-duality along $\hat{R}_1$ to IIB with $N$
fundamental string winding modes, and then an S-duality to IIB with $N$
D1-branes. In this way one arrives at the same description of the DLCQ of
IIA string theory.  This equivalence is of course well known in principle,
and we emphasise it here only because in \cite{csv} the scaling was only
performed after the TS-duality, and even then only somewhat implicitly
(in the definition of the Yang-Mills time variable). In general one
has to perform the scaling rightaway, in conjunction with the boost,
to obtain equivalence with the first prescription (which is rooted firmly
in the Seiberg-Sen derivation of the BFSS matrix theory).

Following this procedure, one finds
that the parameters of the final IIB string theory are related to the $\ell_s,
g_s$ and $R_{11}= \ell_s g_s$ of the original IIA theory by
(we will denote IIB quantities by a prime) 
\be
\label{2blg}
(\ell_s^\prime)^2 = \epsilon (\ell_s)^2\frac{R_{11}}{R}
\qquad\qquad g_s^\prime = \epsilon\frac{R}{R_{11}}\;\;,
\ee
and that the scaled ST-dual metric-dilaton background is given by
\be
\label{dshat}
\begin{aligned}
(ds^\prime)^2 &= \sqrt{g_{11}}\ex{-\phi}\left(
-2d\hat{x}^+d\hat{x}^- + \frac{\ell_s^4}{R^2 g_{11}} (d\hat{x}^1)^2 + \ldots
\right)\equiv \ex{\phi^\prime} d\tilde{s}^2\\
\phi^\prime &= -\phi + \trac{1}{2} \log g_{11}\;\;,
\end{aligned}
\ee
where $\phi$ and $g_{11}$ are functions of $\hat{x}^+ = \tilde{x}^+ = y^+$
(the original lightcone coordinate) and $d\tilde{s}^2$ is the T-dual of the
original scaled IIA metric. Note that this is again a plane wave metric,
written in the almost-Rosen coordinates \eqref{cpwrc}. It is neither
necessary nor convenient to introduce a true Rosen $+$-coordinate 
at this stage since the prefactor of the above metric will in any case drop
out of the DBI action to be discussed below.
We have normalised $\hat{x}^1$ to have
unit radius, $\hat{x}^1 \sim \hat{x}^1 + 2\pi$.
Note that it is due to the fact that we
have implemented the boost by isometries of the metric, and that we have
already performed the scaling, that neither the metric nor the dilaton
has any (undesirable) explicit $\epsilon$-dependence. 

The next step is to look at the Abelian ($N=1$) D1-brane DBI action
in the background \eqref{dshat},
\be
S= 
-\frac{1}{2\pi g_s^\prime\ell_s^{\prime \;2}}
\int d\tau d\sigma\;\ex{-\phi^\prime}
\sqrt{-\det(\del_{\alpha}\hat{x}^{\mu}\del_{\beta}
\hat{x}^{\nu}g^\prime_{\mu\nu}
+ 2\pi \ell_s^{\prime \;2} F_{\alpha\beta})}\;\;,
\ee
and to expand the fields to quadratic order around a suitable 
classical solution of this action. First of all we observe that
\be
\label{ee}
g_s^\prime \ell_s^{\prime\;2} = (\epsilon \ell_s)^2,
\ee
the scaled original
string length squared, and that, correspondingly, the dilaton 
in the DBI action reconverts the ST-dual metric to the T-dual metric 
$d\tilde{s}^2$ implicitly defined in \eqref{dshat}.
We now seek a solution $\hat{x}^{\mu}_c$ of the DBI equations of motion
that describes the groundstate of a simply wrapped
D1-string around the $\hat{x}^1$-direction. Thus we make the 
lightcone gauge ansatz
\be
\hat{x}^+_c = a \tau\quad,\quad
\hat{x}^1_c = b\sigma\quad,\quad \hat{x}^m_c=0\quad,\quad (A_{\alpha})_c =0\;\;.
\ee
The solution for $\hat{x}_c^-$ can then be found
by integrating the lightcone gauge constraint equations
$\del_{\sigma}\hat{x}^-_c =0$ and $\del_\tau \hat{x}^-_c =
b^2\tilde{g}_{11}/2a$.
Compatibility of the classical solution with the periodicity 
of $\hat{x}^1$ and the choice $\sigma \sim \sigma + 2\pi \ell_s$
fixes $b = 1/\ell_s$, and we may as well also choose $a=1$.
Choosing now the gauge $\hat{x}^+=\hat{x}^+_c$, $\hat{x}^1=\hat{x}^1_c$,
the fluctuations are the gauge fields and the fields $(X^1,X^m)$ defined
(with a convenient normalisation) by
\be
\label{fluc}
\hat{x}^-(\tau,\sigma) = \hat{x}^-_c(\tau) + \epsilon \frac{\ell_s}{R}
X^1(\tau,\sigma)
\quad,\quad
\hat{x}^m(\tau,\sigma) = \epsilon X^m(\tau,\sigma)\;\;.
\ee
The fluctuation expansion then becomes an $\epsilon$-expansion (compatible
with the $\epsilon^{-2}$-prefactor arising from \eqref{ee}), and 
to quadratic order in the fluctuations one finds,
after an unenlightning but straightforward calculation,
dropping the field-independent classical action
and a total derivative term, the action
\be
S=\frac{1}{2\pi\ell_s^2}\int d\tau d\sigma [ 
\frac{1}{2}g_{ij}(\tau)(\del_\tau X^i\del_\tau X^j
-\del_\sigma X^i\del_\sigma X^j)
+2\pi^2\ell_s^4 g_s^2 
\ex{2\phi(\tau)}F_{\tau\sigma}^2]\;\;.
\label{max1}
\ee
The final result is  extremely simple. All that enters, after this
sequence of manipulations, are the transverse metric components and
the dilaton of the original IIA configuration, not its T-dualised
or S-dualised cousins.
In particular, the coupling constant of the gauge
theory is set by the original dilaton, 
\be
\label{gym}
g_{YM} \sim \frac{1}{g_s\ell_s} \ex{-\phi}\;\;.
\ee
Moreover, in complete generality the field $X^1$, which began life as a 
fluctuation
of $\hat{x}^-$, in the end plays the role of $x^1$ (which itself had been
gauge fixed). The $\epsilon$-scaling of the fluctuations in \eqref{fluc}
is also natural from this point of view, since it undoes the Penrose scaling 
\eqref{pensc} of the transverse coordinates $\tilde{x}^i$, so that the 
fluctuations are directly related to the coordinates of the original metric
(just as the choice $a=1$ identifies the worldsheet time coordinate with the
original lightcone time coordinate).

It remains to discuss the validity of the truncation of IIB string theory
in the sector with 1 D-brane ($N=1$) to the above action. To that end,
recall first that we had already established in section 2.3
that at the level of gauge
fixed fluctuations (the situation we are dealing with here), the energies
$\d \hat{E}$ 
of the above fluctuation action are finite and equal to the
string theory lightcone energy fluctuations. Thus in order to establish
the decoupling of massive open strings and bulk closed string modes,
we need to compare their energies with the YM energies in the limit
$\epsilon \ra 0$.

In principle, following \cite{csv}, this could be
accomplished by defining a suitable effective string or Planck length
(incorporating the effect of the non-trivial IIB dilaton). However, this
is unnecessary since the metric and dilaton of the final IIB theory are
in any case independent of $\epsilon$ and thus have no bearing on the issue
of decoupling in the DLCQ limit $\epsilon \ra 0$.\footnote{In this respect,
our analysis differs from that of \cite{csv} and, in particular,
subsequent articles that claimed to find 
a much more complicated $\epsilon$-dependence arising from the metric and
dilaton.}
The $\epsilon$-dependence resides only in
the IIB string length and string coupling \eqref{2blg}, both of which
go to zero as $\epsilon \ra 0$. It follows that 
\be
\lim_{\epsilon\ra 0} (\d \hat{E}) \ell_s^\prime 
= \lim_{\epsilon\ra 0} (\d \hat{E}) (g_s^\prime)^{1/4}\ell_s^\prime = 0\;\;,
\ee
which establishes the decoupling of massive open and closed
string modes.

\section{Some Basic Properties of the Matrix String Action for SHPWs}

\subsection{The Plane Wave Matrix String Action in Rosen Coordinates}

It is now reasonable to assume that, in the absence of background fields
other than the metric and the dilaton, the non-Abelian version of the
above action, i.e.\ the decoupled action arising from the sector of
IIB string theory with $N$ would D1-strings, is given by the obvious
non-abelianisation of the above action.\footnote{The presence of other
background fields would complicate matters due to the appearance of Myers
terms etc.} Thus, more or less following the usual matrix string theory conventions
(numerical factors can be changed by various scalings of the fields and
coordinates) and with
the insight that the Yang-Mills coupling constant in this model
is set by the dilaton via \eqref{gym},
the bosonic part of
the matrix string action is (the $X^i=X^i(\tau,\sigma)$ now denote hermitian 
matrix valued fields) 
\begin{multline} 
S=\frac{1}{2\pi\ell_s^2}\int
d\tau d\sigma \Tr \left\{ \frac{1}{2}g_{ij}(\tau)
(D_\tau X^i D_\tau X^j -D_\sigma
X^iD_\sigma X^j)
+ 2\pi^2\ell_s^4 g_s^2 \ex{2\phi(\tau)}F_{\tau\sigma}^2
\right.\\  
\left. 
+\frac{1}{16\pi^2 \ell_s^4g_s^2 } \ex{-2\phi(\tau)} 
g_{ik}(\tau) g_{jl}(\tau) [X^i,X^j]
[X^k,X^l] \right\}\;\;.  
\end{multline}
In terms of the flat
worldsheet metric $\eta_{\alpha\beta}$ this can be written
as 
\begin{multline} 
\label{src1}
S=\frac{1}{2\pi\ell_s^2}\int d^2\sigma \Tr \left\{
-\frac{1}{2}\eta^{\alpha\beta}g_{ij}(\tau)D_\alpha X^i D_\beta X^j -
\frac{1}{4} 4
\pi^2\ell_s^4g_s^2  \ex{2\phi(\tau)}\eta^{\alpha\gamma}\eta^{\beta\delta}
F_{\alpha\beta}F_{\gamma\delta}\right.\\
\left. +\frac{1}{4}
\frac{1}{4\pi^2 \ell_s^4g_s^2 }
\ex{-2\phi(\tau)} g_{ik}(\tau)g_{jl}(\tau)[X^i,X^j] [X^k,X^l] \right\}\;\;.  
\end{multline}
This is the matrix string action for a plane wave in Rosen coordinates,
a convenient coordinate system to start off with since 
the isometry directions required for the reductions, T-duality etc.,
were manifest. However, for many purposes Brinkmann coordinates are
more convenient, and we will
discuss the Brinkmann version of this action below, 
since it also has several advantages over the Rosen coordinate action.

At this point it is appropriate to say a few words about the 
fermionic part of these actions. Apart from the standard kinetic 
term and Yukawa couplings one may ask if there are any other additional
terms arising from the non-trivial target-space metric. Such 
additional terms arise from the spin connection contribution to 
the covariant derivative and have been discussed for D-brane 
actions for instance in \cite{marolfetal}. They are typically of the form
\begin{equation}
\label{connterm}
\bar{\Psi}\Gamma^\alpha\omega_\alpha^{\phantom{\mu}MN}\Gamma_{MN}\Psi
\end{equation}
where $\alpha$ are world-volume indices, 
the $\Psi$ are Majorana spinors and the $\Gamma$ real 
$32\times 32$ gamma matrices (perhaps with a projection on the spinors,
depending on the target space string theory). 

In order to calculate the spin-connection for a plane wave metric in Rosen
coordinates, we introduce the orthonormal frame $E^\pm=dy^\pm, E^a = E^a_i
dy^i$ with $E^a_i$ a vielbein for $g_{ij}(y^+)$. The calculation becomes
particularly simple when one chooses the special (parallel) frame that also
happens to enter in the transformation from Rosen to Brinkmann coordinates
\eqref{rcbc} and which satisfies the symmetry condition \eqref{sc3}. Then 
one finds that the only non-vanishing components of the spin connection
are 
\begin{equation}
\label{spinconn}
\omega^{-a} = \dot{E}_I^a dx^i\;\;.
\end{equation}
In particular, these have no components in the worldsheet directions that
could contribute to \eqref{connterm}. Although we have used the special
symmetric frame to do this calculation, the term that we are considering
(plus the suppressed fermion kinetic term) is covariant under frame
rotations, and the result is thus independent of the choice of frame.

In principle, 
there are also dilatonic contributions to the D-brane action \cite{marolfetal}
which, in the absence of RR fields, take the simple form
\be
\label{phiterm}
\bar{\Psi}\Gamma^{\alpha}\del_{\alpha}\phi\Psi\;\;.
\ee
For Majorana fermions, with $\bar{\Psi}=\Psi^T\Gamma^0$ in a Majorana 
basis, say, this term is also
zero since $\Gamma^0\Gamma^{\alpha}$ is 
symmetric.\footnote{The presence of RR fields would have 
contributed additional projectors to 
\eqref{phiterm}, which could then have given
rise to potentially non-zero contributions
of the dilaton gradient to the D-brane action.} Thus in our case
there are no additional fermionic
contributions to the D-string action arising form either 
the space-time metric or the dilaton.

Since the bosonic part of the matrix string action \eqref{src1} can
be regarded as the dimensional reduction of the 10d Yang-Mills action
in the Rosen coordinate plane wave background to 1+1 dimensions (along
the transverse translational isometries of the plane wave), one might
perhaps have expected the presence of other terms of the same form
as \eqref{connterm} but with a contraction over transverse space-time
indices $i$ rather than the world-volume indices $\alpha$ in the fermionic
part of the matrix string action. While the 
analysis of \cite{marolfetal} suggests that such terms should not appear
in the action, they in any case also turn out to be identically zero
for plane waves. Namely, using \eqref{spinconn} and once again
the symmetry condition
\eqref{sc3}, one finds that the only
possible further contribution to the action is
\begin{equation}
\bar{\Psi}E_A^I\dot{E}_I^A\Gamma^+\Psi\;\;.
\end{equation}
This term, however, is zero for anti-commuting Majorana fermions, for the same
reasons as above.

We close this section with one remark regarding the generality of the above
action. Strictly speaking, 
we have derived this action only for singular homogeneous plane waves
since in that case we could implement almost literally the accepted
flat space Seiberg-Sen DLCQ prescription (in its CSV variant). However, 
as it stands, this action makes sense for any plane wave metric - dilaton
system. If one can argue that, in order to correctly implement the DLCQ
in curved backgrounds, one only needs to consider the combined boost-scaling
Penrose transformation \eqref{pensc}, as e.g.\ the arguments in \cite{shomer}
suggest, then the entire derivation of the matrix string action, including
the decoupling arguments, still goes through, and one would then have
estabished the validity of the above matrix string action for any plane wave.

\subsection{Rosen vs Brinkmann Form of the Matrix String Action}

Here we discuss the Brinkmann coordinate (BC) counterpart of the Rosen
coordinate (RC) matrix Yang-Mills action \eqref{src1}, which we now write in
slightly simplified form as
\begin{multline}
\label{src}
S_{RC} = \int d^2\sigma\Tr\left( -\trac{1}{4}g^{-2}_{YM}
\eta^{\alpha\gamma}\eta^{\beta\delta}F_{\alpha\beta}F_{\gamma\delta}
-\trac{1}{2}\eta^{\alpha\beta}g_{ij}(\tau)D_{\alpha}X^i D_{\beta}X^j
\right. \\ \left. 
+ \trac{1}{4}g^2_{YM} g_{ik}(\tau)g_{jl}(\tau)[X^i,X^j][X^k,X^l] \right)
\;\;.
\end{multline}
Now in Brinkmann coordinates a plane wave takes the form
\eqref{pwbc}, 
\be
-2dx^+dx^- + g_{ij}(x^+)dx^i dx^j = 
 -2dz^+dz^- + A_{ab}(z^+)z^az^b (dz^+)^2 + \d_{ab}dz^a dz^b
\ee
and since typically (e.g.\ in the lightcone gauge point particle or string
actions) $A_{ab}$ turns into a mass term for the fields, 
as a potential Brinkmann counterpart of this action, we consider
the action
\begin{multline}
\label{sbc}
S_{BC} = \int d^2\sigma\Tr\left( -\trac{1}{4}g^{-2}_{YM}
\eta^{\alpha\gamma}\eta^{\beta\delta}F_{\alpha\beta}F_{\gamma\delta}
-\trac{1}{2}\eta^{\alpha\beta}\d_{ab}D_{\alpha}Z^a D_{\beta}Z^b\right. \\
\left. + \trac{1}{4}g^2_{YM} \d_{ac}\d_{bd}[Z^a,Z^b][Z^c,Z^d] +
\trac{1}{2}A_{ab}(\tau)Z^aZ^b \right)\;\;.
\end{multline}
On the face of it, these two classes of actions appear to be rather
different, with \eqref{src} having non-standard time-depenent kinetic
terms and quartic couplings for the scalar fields, described by the
$g_{ij}(t)$, while \eqref{sbc} has standard kinetic and quartic terms
but time-dependent mass terms for the scalars (with $A_{ab}(t)$ minus
the mass-squared matrix).
Nevertheless, we claim that these two types of
Yang-Mills actions are
simply related by a certain linear field redefinition 
$X^i=E^i_{\;a}(\tau)Z^a$  of the scalar fields, 
\be
\label{srcsbc}
S_{RC}[A_{\alpha},X^i=E^i_{\;a}Z^a] = S_{BC}[A_{\alpha},Z^a]\;\;.
\ee
To see this, recall 
first of all the coordinate transformation \eqref{rcbc} 
between Rosen and Brinkmann coordinates, in particular the part
$x^i= {E}^i_{\;a}z^a$.
We are thus led to consider the linear field transformation
\be
X^i(\tau,\sigma) = E^i_{\;a}(\tau) Z^a(\tau,\sigma)
\label{XEZ}
\ee
of the scalar fields (matrix-valued coordinates) $X^i$ and $Z^a$, where
$E^i_{\;a}$ is a vielbein for the time-dependent metric
$g_{ij}(\tau)$ on the scalar field space satisfying the symmetry condition
(\ref{sc3}).
Substituting \eqref{XEZ} into the RC 
Lagrangian $L_{RC}$, one can
now verify that one indeed
obtains the BC Lagrangian $L_{BC}$ up to a total time-derivative
(related to the shift of $x^-$ in \eqref{rcbc}).
We explain this calculation in somewhat
more detail in \cite{blrcbc}, where we also show that this kind of
argument extends to the plane wave counterparts of the recently
proposed multiple M2-brane actions \cite{bl,gu,lor3} 
based on 3-algebras rather
than Lie algebras. 
Here we just want to point out that a crucial role in this calculation
is played by the
symmetry condition \eqref{sc3} (and, of course, by the gauge invariance of 
the action). Not only is this condition responsible
for several cancellations that are akin to those that already occur in
the transformation of a plane wave metric from Rosen to Brinkmann coordinates.
It cooperatively also
serves to eliminate some terms of genuinely non-Abelian origin, such as
\be
g_{ij}(t)E^i_{\;a} \dot{E}^j_{\;b} \Tr [A_t,Z^a] Z^b
= g_{ij}(t)E^i_{\;a} \dot{E}^j_{\;b} \Tr A_t[Z^a, Z^b] =0
\ee
arising from the scalar kinetic terms.
The above equivalence \eqref{srcsbc} is also valid for a time-dependent
coupling constant $g_{YM}(\tau)$, since the total time-derivative arises only
from the dilaton-independent scalar kinetic term. It also extends to the
fermionic terms in the action in a rather trivial way, since the (Yukawa)
coupling between the fermions and the scalars $X^i$ is purely algebraic.

The main advantage of the Brinkmann form \eqref{sbc} of the action is
that the scalar fields have standard kinetic terms. This implies that it
is legitimate and
meaningful to look at the potential terms to deduce some properties
of the classical and quantum theories. In particular, whether one is in
an Abelian or non-Abelian phase of the theory can be reliably read off
from the behaviour of the dilaton. For example, when the dilaton blows
up at the singularity (a strong coupling singularity in the sense of the
analysis in appendix B.3), the Yang-Mills coupling constant \eqref{gym}
is small and one is in a genuinely non-Abelian phase of the theory, just
as in the CSV model, where the matrix coordinates are non-commutative. 
By contrast,
conclusions based solely on the analysis of the quartic potential term
in the RC action \eqref{src}, or the attempt to read off something like an effective string tension from 
the RC kinetic term, are bound to be misleading at best.

Furthermore, the mass terms in the BC action contain direct invariant
geometric information about the space-time, since they arise from the
components of the Riemann tensor in Brinkmann coordinates. In particular,
we will see below that they faithfully encode the information whether
one is dealing with a strong or weak coupling singularity (in the sense
of the analysis in appendix B.3), something that is not at all manifest
in the RC action which also exhibits spurious coordinate singularities.

\subsection{Absorbing the Coupling Constant into the Worldsheet Metric}

It is obvious, and a basic property of 2-dimensional gauge theories,
that the dilaton / Yang-Mills coupling constant can, in
either the Rosen or the Brinkmann form of the Yang-Mills action, in
principle always 
be absorbed into a non-trivial worldsheet metric via
\be
\label{hab}
h_{\alpha\beta} = \ex{-2\phi}\eta_{\alpha\beta}\;\;,
\ee
since one then has
\be
\sqrt{h} = \ex{-2\phi} \qquad
\sqrt{h}h^{\alpha\beta}  = \eta^{\alpha\beta} \qquad
\sqrt{h}h^{\alpha\gamma}h^{\beta\delta}   = 
\ex{2\phi} \eta^{\alpha\gamma}\eta^{\beta\delta} \;\;,
\ee
which are precisely the pre-factors of the quartic, scalar 
kinetic and $F^2$ terms respectively.
Once one has absorbed the dilaton into the worldsheet metric (in view
of the considerations below it is not clear if one really wants to do
this in general), the only time-dependence remaining in the Brinkmann
coordinate matrix string action is in the mass terms $A_{ab}(\tau)$.
The Rosen coordinate matrix string theory action, on the other hand,
still has explicit time-dependence arising from the metric coefficients
$g_{ij}(\tau)$.  Such a remaining time-dependence can never be absorbed
by a further (conformal) redefinition of the worldsheet metric since
the combination $\sqrt{h}h^{\alpha\beta}$ is conformally invariant so
that the time-dependence in the kinetic term $g_{ij}DX^iDX^j$ can not
be eliminated in this way.

In the CSV model (flat metric with a linear dilaton), \eqref{hab}
resulted in a useful
alternative description of the theory, heavily made use of e.g.\
in \cite{csv2}.
Instead of a Yang-Mills theory
with a time-dependent coupling constant on a cylindrical worldsheet with
the trivial metric one then has a Yang-Mills theory
with a time-independent coupling constant on a cylindrical worldsheet
with a non-trivial time-dependent metric. With $\phi = -Q\tau$, 
the line-element is
\be
ds^2 = \ex{2Q\tau}(-d\tau^2 + d\sigma^2)\;\;.
\ee
While this metric is locally flat, the periodicity of $\sigma$
results in the worldsheet being the Milne orbifold \cite{csv}.

For the backgrounds with $b\neq -1$, on the other hand, we have
$\exp{2\phi(\tau)} = \tau^{3b/(b+1)}$.
Thus the rescaled worldsheet metric has the form
\be
ds^2 = \tau^{2\gamma}(-d\tau^2 + d\sigma^2)
\ee
where $2\gamma = -3b/(b+1)$. The Einstein-dilaton equations imply that 
$\gamma \geq -1$ \eqref{3b2}.
Now this metric is singular even prior to the
periodic identification of $\sigma$ unless either, trivially
$\gamma=0$ (i.e.\ $b=0$, a constant dilaton), 
or $\gamma =-1$. 
Indeed, when $\gamma\neq -1$, the metric has a curvature singularity
at $\tau=0$, as can be seen 
by calculating e.g.\ the Ricci scalar
$R = -2\gamma\tau^{-(2\gamma + 2)}$.
For $\gamma=-1$, on the other hand, $R$ is constant and 
with $T=\log\tau$ one has
\be
ds^2 = \tau^{-2}(-d\tau^2 + d\sigma^2) = -dT^2 + \ex{-2T}d\sigma^2\;\;.
\ee
This is just the $(1+1)$-dimensional de Sitter (dS) metric, written in 
coordinates that cover half of the entire dS space-time. 
$\gamma=-1$ corresponds to $3b/(b+1)=2$, i.e.\ $b=2$.  This is the dual
background (under the $b\ra 1-b$ isometry) of the CSV solution (see the
remark after \eqref{bc}), and thus the dS worldsheet arises in
the dual reduction of the CSV M-theory background \eqref{acsvm}.

Periodicity of $\sigma$ means that we are considering here dS space-time
with toroidal (rather: circular) spatial sections. While on the face of
it this appears to be an innocuous modification of the dS metric, this
space-time is, in spite of apparently being non-singular, actually known
to be timelike geodesically incomplete \cite{galloway,brett} (i.e.\ it
contains inextendible timelike geodesics of finite length).  Note that
this worldsheet geodesic incompleteness appears at $T\ra-\infty$,
i.e.\ at the location of the space-time singularity.  Nevertheless,
this worldsheet structure may be more tractable than the (genuinely
singular) worldsheets that arise for $\gamma\neq -1$.  In particular,
here the point of incompleteness is ``inaccessible'' in the sense that
an observer who wants to reach it in finite proper time needs to wrap
around the circle an infinite number of times.

\subsection{Tachyons and Strong String Coupling Singularities}

Let us take a closer look at the scalar sector of the BC action \eqref{sbc}.
The information about the metric resides solely in the mass matrix
\be
\mu^2_{ab}(\tau) = -A_{ab}(\tau) \;\;.
\ee
For the singular homogeneous plane wave backgrounds
of Appendix B one has \eqref{bn1}
\be
\mu^2_{ab}(\tau) = \mu^2_a(\tau)\d_{ab} = -m_a(m_a-1) \tau^{-2}
\d_{ab}\;\;.
\ee
Now by the Einstein-dilaton equation \eqref{ede2}, 
\be
\sum_a m_a(m_a-1) = -\frac{3b}{b+1}\;\;,
\ee
the parameters $m_a$ are related to the parameter $b$ determining the
dilaton $\phi$. The sign of $b$ in turn determines (appendix B.3)
whether $\exp \phi$
blows up at the singularity (strong coupling singularity, $b<0$) or goes to
zero there (weak coupling singularity, $b>0$). In particular, if all the 
mass-squares $\mu_a^2$ of the scalars are positive, necessarily $b$ is
positive, and one is dealing with a weak coupling singularity. Conversely,
therefore, whenever one is dealing with a strong coupling singularity, 
at least one of the scalars is tachyonic (and the sum over all the $\mu_a^2$
is negative). Thus this is the way a strong
coupling singularity manifests itself in the BC matrix string action.

The derivation of the model, in particular the decoupling analysis
of section 2.4, suggests that the matrix string action gives a valid
description of the string theory at least for all $\tau >0$ and even
when $\tau\ra 0$. Thus the presence of these tachyonic mass terms,
which can in principle occur both for strong and for weak coupling
singularities, should not all by itself be indicative of a pathology of
the model.\footnote{In this context it may be worth pointing out that
in a related setting, namely the modelling of cosmological
singularities via AdS/CFT, it was also found to be necessary to introduce
a potential that is unbounded from below \cite{cht}. More recently, tachyonic mass terms
of the above $\tau^{-2}$-type have also been shown to arise naturally
in the cosmological AdS/CFT context \cite{addnnt}.} In particular,
in the non-Abelian phase of the theory the tachyonic mass
terms can potentially be stabilised by the quartic potential, perhaps
indicating the existence of some new and interesting
non-perturbative physics.
One might e.g.\ like to see if there is a qualitatively
different behaviour for strong ($\sum_a \mu_a^2 <0$) vs weak ($\sum_a
\mu_a^2 >0$) coupling singularities. It may also be of interest to
study the implications of the classical scale invariance of these models,
manifested e.g.\ in the characteristic $\tau^{-2}$-dependence of the
mass terms, in the quantum theory.

\subsection*{Acknowledgements}

We are very grateful to Denis Frank, Giuseppe Milanesi and Sebastian Weiss
for numerous helpful discussions during the long gestation-period of
this project, and for their collaboration on related matters. This work has
been supported by the Swiss National Science Foundation and by the EU
under contract MRTN-CT-2004-005104.

\appendix
 
\section{Plane Wave Geometry: Synopsis}

There are two standard coordinate systems for plane wave metrics, each
with its own advantages. In \textit{Rosen coordinates}, the metric 
takes the form
\be
d{s}^2 = g_{\mu\nu}dy^\mu dy^\nu = -2 dy^+ dy^- + {g}_{ij}(y^+)dy^i dy^j \;\;.
\label{pwrc}
\ee
In Rosen coordinates it is manifest that any metric conformal to
a plane wave metric, with the conformal factor depending only on $y^+$,
\be
ds^2 = f(y^+)( -2 dy^+ dy^- + {g}_{ij}(y^+)dy^i dy^j)\;\;, 
\label{cpwrc}
\ee
or,  equivalently, a metric of the type
\be
ds^2 = -2 f(y^+)dy^+ dy^- + {g}_{ij}(y^+)dy^i dy^j\;\;, 
\label{cpwrc2}
\ee
is again a plane wave metric, as can be sen by defining the new Rosen
coordinate (affine parameter) $\tilde{y}^+$ by $d\tilde{y}^+ = f(y^+) dy^+$.

The metric \eqref{pwrc} 
has the manifest commuting translational Killing vectors $Z=\del_{y^-}$ and 
$Q_{(i)}=\del_{y^i}$. In addition, any plane wave metric has 
the ``hidden'' dual commuting translational Killing vectors 
\be
\label{api}
P^{(i)} = y^i \del_{y^-} + h^{ik} \del_{y^k}\;\;,
\ee
where
\be
\label{hik}
h^{ik}(y^+) = \int^{y^+} du\; g^{ik}(u)\;\;,
\ee
which extend the Abelian isometry algebra generated by $Q_{(i)}$ and
$Z$ to the Heisenberg algebra $[Q_{(i)},P^{(k)}]=\d_i^{\;k}Z$
with central element $Z$.

Rosen coordinates are not the coordinate system in which plane waves
are usually and most conveniently discussed, among other reasons
because typically in Rosen coordinates the metric
exhibits spurious coordinate singularities.
The plane wave metric in \textit{Brinkmann coordinates} is
\be
\label{pwbc}
d{s}^2 = g_{\mu\nu}dz^\mu dz^\nu =
 -2dz^+dz^- + A_{ab}(z^+)z^az^b (dz^+)^2 + \d_{ab}dz^a dz^b\;\;.
\ee
Brinkmann coordinates are \textit{Fermi
coordinates} adapted to the null geodesic $(z^+=\tau,z^-=0,z^a=0)$
\cite{bfw2}. In particular, Brinkmann coordinates are, like Riemann 
coordinates, a direct measure of the invariantly defined geodesic distance 
in space-time. 

Moreover, and related to this, in Brinkmann coordinates, the curvature
of the plane wave is related purely algebraically to the mass/frequency
term $A_{ab}(u)$ of the metric, which trivialises the task of calculating
the curvature of a plane wave. Specifically, the only non-vanishing
components of the Riemann tensor and Ricci tensor are
\be
\label{rpwbc}
R_{+a+b}(z^+) = - A_{ab}(z^+) \qquad\qquad
R_{++}(z^+) = - \d^{ab}A_{ab}(z^+) \;\;,
\ee
and the Ricci scalar is zero. Thus the metric is flat iff $A_{ab}=0$ and
the vacuum Einstein equations reduce to the simple algebraic condition on
$A_{ab}$ (regardless of its $z^+$-dependence) that it be traceless. The
number of degrees of freedom of this traceless matrix $A_{ab}(z^+)$
are those of a transverse traceless symmetric tensor (a.k.a.\ a graviton).

The two classes of metrics described by (\ref{pwrc}) and (\ref{pwbc})
are equivalent: every metric of the form (\ref{pwrc}) can be brought to
the form (\ref{pwbc}), and conversely every metric of the type (\ref{pwbc})
can be written, in more than one way, as in (\ref{pwrc}).
The coordinate transformation relating \eqref{pwrc} and \eqref{pwbc} is
\be
(y^+,y^-,y^i) =(z^+,z^- - \trac{1}{2}\dot{{E}}_{ai}{E}^i_{\;b}z^a z^b,
{E}^i_{\;a}z^a)\;\;,
\label{rcbc}
\ee
where $E^i_{\;a}=E^i_{\;a}(y^+)$ is a vielbein for ${g}_{ij}(y^+)$, 
$E^i_{\;a}E^j_{\;b}{g}_{ij}=\d_{ab}$,
subject to the symmetry condition
\be
\dot{{E}}_{ai}{E}^i_{\;b} = \dot{{E}}_{bi}{E}^i_{\;a}
\label{sc3}
\ee
(which can be interpreted \cite{bbop1,bbop2}
as the condition that the frame $E^i_{\;a}$ is
parallel transported along the congruence of null geodesics defined by the
Rosen coordinates). The relation between $g_{ij}(y^+)$ and $A_{ab}(z^+)$ 
can be succinctly written as \cite{mmhom} (recall that $z^+=y^+$)
\be
A_{ab}(z^+)= \ddot{{E}}_{ai}(z^+) {E}^i_{\;b}(z^+)\;\;.
\ee
The above relations simplify considerably for diagonal metrics,
$g_{ij}(y^+)=g_i(y^+)^2\d_{ij}$,
for which one simply has
\be
A_{ab}(z^+) = \frac{\ddot{g}_a(z^+)}{g_a(z^+)}\d_{ab}\;\;.
\ee
In particular, for Rosen coordinate metrics of power-law type, 
\be
\label{arcpl}
ds^2 = - 2dy^+ dy^- + \sum_{i} (y^+)^{2m_i}(dy^i)^2\;\;,
\ee
the metric in Brinkmann coordinates is
\be
\label{abcpl}
ds^2 = -2dz^+dz^- + \sum_a \frac{m_a(m_a-1)}{(z^+)^2}(z^a)^2 (dz^+)^2 
+ \sum_a (dz^a)^2\;\;,
\ee
and its Ricci tensor is
\be
\label{ariccipl}
R_{++}(z^+) = \sum_a m_a(m_a-1) (z^+)^{-2}\;\;.
\ee
Since $m_a(m_a-1)$ is invariant under $m_a \ra 1-m_a$, \eqref{abcpl} shows that
Rosen coordinate metrics with $m_i$ and $m_i\ra 1-m_i$ are isometric. In
particular, any metric with all $m_i=0,1$ is flat. 

It is evident e.g.\ from \eqref{ariccipl} that these plane waves are singular
at $z^+=0$ and, since $z^+$ can play the role of an affine geodesic
parameter, that this singularity is at finite affine distance, so that these
metrics are geodesically incomplete.
Moreover, these power-law metrics and their Brinkmann coordinate counterparts 
have the
special property that they are \textit{scale-invariant}, i.e.\ invariant
under scalings of the coordinate (affine parameter)
$y^+$ or $z^+$. This is evident for \eqref{abcpl}, which is invariant under 
\be
(z^+,z^-) \ra (\lambda z^+,\lambda^{-1}z^-)\;\;.
\ee
Thus these metrics have the additional Killing vector
$X = z^+\del_{z^+} - z^-\del_{z^-}$.
The corresponding isometry in Rosen coordinates \eqref{arcpl} is
\be
(y^+,y^-,y^i) \ra (\lambda y^+,\lambda^{-1}y^-,\la^{-m_i}y^i)\;\;.
\ee
See \cite{prt} and \cite{mmhom} for a systematic discussion 
and other properties of these singular homogeneous plane waves.

\section{A Class of Plane Wave -- Null Dilaton Big Bang Backgrounds}

\subsection{M$\ra$IIA Reduction of Singular Homogeneous 
Plane Wave Backgrounds}

Using the standard relation
\be
ds_{11}^2 = \ex{-2\phi/3}ds_{st}^2 + \ex{4\phi/3}dy^2
\ee
between M-theory and IIA string frame backgrounds, one sees that
the CSV \cite{csv} configuration
\be
\label{csv}
ds_{st}^2 = -2dy^+ dy^- + \d_{ij}dy^i dy^j \qquad\qquad
                  \ex{2\phi} = \ex{-3y^+}\;\;,
\ee
Minkowski space with a linear dilaton, lifts to the M-theory 
plane wave metric 
\be
\label{acsvm}
ds_{11}^2 = -2dudv + u\d_{ij}dy^idy^j + u^{-2} (dy)^2\;\;,
\ee
where $y^+ = \log u$ (and $y^-=v$). We will now turn this around and
consider the reduction of more general 
11d plane wave metrics (in Rosen coordinates)
\be
ds_{11}^2 = -2dudv + G_{ij}(u)dy^i dy^j + c(u)dy^2
\ee
along $y$. Then 
one obtains the IIA string frame metric + null dilaton configuration
\be
\begin{aligned}
ds_{st}^2&=c^{1/2}(u)(-2dudv + G_{ij}(u)dy^idy^j) \\
\ex{2\phi(u)} &= c(u)^{3/2}\;\;.
\end{aligned}
\ee
The ten-dimensional string and Einstein frame metrics are also plane waves.
In the string frame, the standard Rosen form is obtained by introducing the 
null coordinate (affine parameter) $y^+$,
\be
\label{dyu}
dy^+ = c^{1/2}(u)du\;\;,
\ee
in terms of which (and $y^-=v)$ the metric takes the standard Rosen form
\eqref{pwrc} with $g_{ij}(y^+) = (c^{1/2}G_{ij})(u(y^+))$.
The Einstein frame metric $ds_{e}^2 = \exp{(-\phi/2)}ds_{st}^2$
is also manifestly a plane wave, written in the almost-Rosen form \eqref{cpwrc}.

Let us now concentrate on the 11d singular homogeneous plane waves
of the power-law form 
\be
\label{rcpl2}
\begin{aligned}
ds_{11}^2 &= -2dudv + \sum_i u^{2n_i}(dy^i)^2 + u^{2b}(dy)^2\\
&= - 2dz^+ dz^- + \sum_a \frac{n_a(n_a-1)}{(z^+)^2}(z^a)^2 (dz^+)^2 +
\frac{b(b-1)}{(z^+)^2}z^2 (dz^+)^2 + \sum_a (dz^a)^2 + (dz)^2 \\
\end{aligned}
\ee
The relation $dy^+ = u^b du$ \eqref{dyu} integrates to
$y^+ = u^{b+1}/(b+1)$ for $ b \neq -1$ and to  $y^+=\log u$ for 
$b=-1$. Thus the dilaton behaves as 
\be
\label{dilaton}
\ex{2\phi(u)}= u^{3b} = \left\{\begin{array}{cc} [(b+1)y^+]^{3b/(b+1)} & b
\neq -1 \\ \ex{3b y^+}=\ex{-3y^+} & b = -1
\end{array}\right.
\ee
While in general one obviously always finds a null dilaton
in 10 dimensions, a linear dilaton, as in the CSV model \eqref{csv},
arises only for the special value $b=-1$ of the parameter $b$.
By a suitable scaling of the coordinates, one can put the $b\neq -1$ 
IIA backgrounds into the normalised form (\ref{arcpl},\ref{abcpl})
\be
\begin{aligned}
\label{bn1}
ds_{st}^2 &= -2 dy^+ dy^- + \sum_{i} (y^+)^{2m_i} (dy^i)^2\\
&= -2dz^+dz^- + \sum_a \frac{m_a(m_a-1)}{(z^+)^2}(z^a)^2 (dz^+)^2 
+ \sum_a (dz^a)^2\\
\ex{2\phi} &= (y^+)^{3b/(b+1)}\;\;,
\end{aligned}
\ee
with
\be
\label{mdef}
2m_i = \frac{2n_i+b}{b+1}\;\;.
\ee
For $b=-1$, on the other hand, one has
\be
\begin{aligned}
\label{b1}
ds_{st}^2 &= -2 dy^+ dy^- + \sum_{i} e^{(2n_i -1)y^+} (dy^i)^2 \\ 
&= -2dz^+ dz^- + \sum_a (2n_a-1)^2 (z^a)^2 (dz^+)^2 + \sum_a (dz^a)^2
\end{aligned}
\ee
with a linear dilaton.
This has the standard form of a metric of a non-singular
symmetric plane wave (constant $A_{ab}$).

\subsection{Equations of Motion}

The 11d vacuum Einstein equations for the singular homogeneous plane
wave \eqref{rcpl2} reduce to the algebraic condition
\be
\label{nnbb}
\sum_{a} n_a(n_a-1) + b(b-1) = 0\;\;.
\ee
In terms of the IIA parameters $m_i$ \eqref{mdef} for $b \neq -1$, 
this equation can be written as
\be
\label{ede2}
\sum_i m_i(m_i-1) = -\frac{3b}{b+1}\;\;,
\ee
which implies
\be
\label{3b2}
\frac{3b}{b+1}\leq 2\;\;.
\ee
The algebraic constraint \eqref{ede2}
can be recognised as the Einstein-dilaton equation
\be
\label{ede}
R_{++}(z^+) = -2\del_+\del_+ \phi(z^+)
\ee
for the dilaton
\be
\label{dsol}
\phi(z^+) = \frac{3b}{2(b+1)}\log z^+\;\;.
\ee
Since all terms in \eqref{nnbb}
are positive when the parameters are sufficiently large
(positive or negative), this strongly constrains their allowed range.
A useful way of writing this equation is
\be
\label{eom1}
(b-2)(b+1) = - \sum_{a} (n_{a}-1/2)^2\;\;.
\ee
Since the right hand side is non-positive, this leads to the
constraint
\be
\label{bc}
-1 \leq b \leq 2\;\;.
\ee
Thus the linear dilaton case $b=-1$ lies at the boundary of the allowed
parameter range, and the only solution with $b=-1$ is the (lifted)
CSV solution \eqref{acsvm} with $n_a=1/2$. The solution with $b=2$ is, 
in a sense dual to the CSV background (the 11d metrics with $b=-1$ and
$b=2$ are isometric, but the reduction to 10d is performed either
along $\del_y$ or along the dual isometry direction).

\subsection{Singularity Structure and 
Behaviour of the Dilaton at the Singularity}

We now want to analyse the behaviour of the dilaton at the singularity of the 
plane wave metrics for $b\neq -1$, and begin with a brief review of the
situation for the $b=-1$ CSV background.
As a symmetric plane wave, the $b=-1$ metric (\ref{b1}) is completely
non-singular. It is isometric to the flat metric iff $2n_a-1=0$.
This is precisely the CSV background (\ref{csv},\ref{acsvm}), and the
only solution to the vacuum Einstein equations for $b=-1$.

Even though the string frame metric is flat, the IIA background as a
whole should be considered to be singular \cite{csv}, either because of
the dilaton $\sim \exp{-3y^+}$, which is singular as $y^+ \ra -\infty$,
or because of the behaviour of the metric in the Einstein frame. This is
compatible with the fact that the M-theory lift \eqref{acsvm} of the CSV
background is itself a singular homogeneous plane wave with a singularity
at $u=0$ ($y^+=\log u$). 
Thus the singularity
arises at strong string coupling and therefore, in the matrix string
setting, at weak gauge coupling. Far from the singularity, on the other
hand, the string coupling goes to zero.
Since $y^+$ can be identified with the relevant
affine geodesic parameter, the singularity is located at infinite
geodesic distance in the string frame metric.  This is in contrast to
what happens for the 11-dimensional lift of the CSV metric \eqref{acsvm}
(the singularity occurs at the finite value $u=0$ of the affine parameter
$u$) or in the Einstein frame.  

The  $b\neq - 1$ metrics (\ref{bn1}) have a singularity at $y^+=0$
unless the metric is isometric to the flat metric, which is the case
iff $m_a(m_a-1)=0$, i.e.\ $m_a=0$ or $m_a=1$, requiring also a constant 
dilaton $b=0$.
The relation between the 11-dimensional and 10-dimensional
string frame affine paramters $u$ and $y^+$ is 
\be
u \sim (y^+)^{1/(b+1)}
\ee
and the dilaton is
\be
\ex{2\phi} \sim (y^+)^{3b/(b+1)} \;\;.
\ee
Thus for $ b+1 > 0$ (\ref{bc}), a condition implied by the Einstein
equations, $u\ra 0$ corresponds to $y^+\ra 0$, and thus
the singularity is always at finite geodesic distance,
even in the string frame, in contrast to what happens in the CSV background.

Since $b+1 >0$, it is evident that the behaviour of the dilaton at the
singularity is determined by the sign of $b$,
\be
\begin{aligned}
\text{strong coupling singularity:}& \qquad -1 < b < 0 \\
\text{weak coupling singularity:}& \qquad \phantom{-} 0 < b \leq 2\;\;.
\end{aligned}
\ee
Finally, note that in the strongly coupled range
the string coupling goes to zero at infinity, i.e.\
as $z^+\ra\infty$, just as in the CSV model. 
In the weakly coupled range, on the other hand, the 
string coupling would blow up there. This can be remedied by noting \cite{prt}
that the general solution of the IIA Einstein-dilaton equations \eqref{ede},
also includes a linear term in the dilaton solution \eqref{dsol},
\be
\phi(z^+) = \frac{3b}{2(b+1)}\log z^+ - c z^+\;\;.
\ee
For $c>0$ and $b>0$ this has the effect that the string coupling now
tends to zero both at the singularity $z^+=0$ and for $z^+ \ra \infty$.
This is the case analysed from a string theory point of view in \cite{prt}.
Since the metric-dilaton background for $c\neq 0$ does not arise from (or
lift to) a singular homogeneous plane wave in 11 dimensions, and 
since, in the spirit of \cite{bbop1,bbop2}, we regard the singular homogeneous
plane wave metrics not as genuine comsological toy-models but as a
near-singularity approximation of a singular space-time (in particular, we do
not trust / take seriously the metric as $z^+\ra\infty$), 
we will only consider the solutions with $c=0$.

\rnc{\Large}{\normalsize}


\begin{thebibliography}{00}
\addcontentsline{toc}{section}{References}
\frenchspacing
\begin{small}
\addtolength{\itemsep}{-4pt}
\nc{\ct}[1]{\textit{#1}}

\bibitem{berkooz} M. Berkooz, D. Reichmann, \textit{A Short Review of Time
Dependent Solutions and Space-like Singularities in String Theory},
\texttt{arXiv:0705.2146v1 [hep-th]}.

\bibitem{bfss} T. Banks, W. Fischler, S. Shenker and L. Susskind,
\textit{M theory as a matrix model: A Conjecture}, 
Phys. Rev. D55 (1997) 5112, \texttt{arXiv:hep-th/9610043}. 

\bibitem{susskind} L. Susskind, \textit{Another conjecture about 
M(atrix) theory}, \texttt{arXiv:hep-th/9704080}.

\bibitem{motletc} L. Motl, \textit{Proposals on nonperturbative 
superstring interactions}, \texttt{arXiv:hep-th/9701025};
T.Banks and N. Seiberg, \textit{Strings from matrices}, 
Nucl. Phys. B497 (1997) 41, \texttt{arXiv:hep-th/9702187}.

\bibitem{dvv} R. Dijkgraaf, E.P. Verlinde and H.L. Verlinde, 
\textit{Matrix string theory}, Nucl. Phys. B500 (1997) 43,
\texttt{arXiv:hep-th/9703030}.

\bibitem{vrwt} 
W. Taylor, M. van Raamsdonk, 
\textit{Multiple D0-branes in Weakly Curved Backgrounds},
Nucl.Phys. B558 (1999) 63-95, 
\texttt{arXiv:hep-th/9904095v2};
W. Taylor, M. van Raamsdonk, 
\textit{Multiple Dp-branes in Weak Background Fields},
Nucl.Phys. B573 (2000) 703-734,
\texttt{arXiv:hep-th/9910052v1}.

\bibitem{schiappa} R. Schiappa, \textit{Matrix Strings in Weakly Curved
Background Fields}, Nucl.Phys. B608 (2001) 3-50,
\texttt{arXiv:hep-th/0005145v2}.

\bibitem{csv} B. Craps, S. Sethi and  E.P. Verlinde, 
\textit{A Matrix Big Bang}, JHEP 0510:005, (2005),
\texttt{arXiv:hep-th/0506180}. 

\bibitem{csv2} B. Craps, A. Rajaraman and S. Sethi, \textit{Effective 
Dynamics of the Matrix Big Bang}, Phys. Rev. D73 (2006) 106005,
\texttt{arXiv:hep-th/0601062}.

\bibitem{craps} B. Craps, \textit{Big Bang Models in String Theory},
Class. Quant. Grav. 23 (2006) S849, \texttt{arXiv:hep-th/0605199}.

\bibitem{null1} D. Robbins and S. Sethi, \textit{A Matrix Model for the 
Null-Brane}, JHEP 0602 (2006) 052, \texttt{arXiv:hep-th/0509204}.

\bibitem{null2} E. Martinec, D. Robbins and S. Sethi, \textit{Toward the 
End of Time}, JHEP 0608 (2006) 025, \texttt{arXiv:hep-th/0603104}.

\bibitem{brunch} J. Bedford, C. Papageorgakis, D. Rodriguez-Gomez and J. Ward, 
\textit{Matrix Big Brunch}, \texttt{arXiv:hep-th/0702093}.

\bibitem{miaoli} M. Li, \textit{A Class of Cosmological Matrix Models},
Phys. Lett. B626 (2005) 202, \texttt{arXiv:hep-th/0506260}; 
M. Li and W. Song, \textit{Shock Waves and Cosmological Matrix Models},
JHEP 0510 (2005) 073, \texttt{arXiv:hep-th/0507185}. 

\bibitem{chen} B. Chen, \textit{The Time-dependent Supersymmetric 
Configurations in M-theory and Matrix Models}, Phys. Lett. B632 (2006) 393,
\texttt{arXiv:hep-th/0508191}; Hong-Zhi Chen and Bin Chen, \textit{Matrix Model 
in a Class of Time Dependent Supersymmetric Backgrounds}, 
Phys. Lett. B638 (2006) 74, \texttt{arXiv:hep-th/0603147}.

\bibitem{dm} S.R. Das and J. Michelson, \textit{pp wave big bangs: 
Matrix strings and shrinking fuzzy spheres}, 
Phys. Rev. D72 (2005) 086005, \texttt{arXiv:hep-th/0508068}; 
S.R. Das and J. Michelson, \textit{Matrix membrane big bangs and 
D-brane production}, Phys. Rev. D73 (2006) 126006, 
\texttt{arXiv:hep-th/0602099}.

\bibitem{ohta2} T. Ishino and N. Ohta, \textit{Matrix String Description 
of Cosmic Singularities in a Class of Time-dependent Solutions}, 
Phys. Lett. B638 (2006) 105, \texttt{arXiv:hep-th/0603215}.

\bibitem{seiberg} N. Seiberg, \textit{Why is the Matrix Model Correct?},
Phys. Rev. Lett. 79 (1997) 3577, \texttt{arXiv:hep-th/9710009}.

\bibitem{sen} A. Sen, \textit{D0 Branes on $T^n$ and Matrix Theory}, 
Adv. Theor. Math. Phys. 2 (1998) 51, \texttt{arXiv:hep-th/9709220}.

\bibitem{sen2} A. Sen, \textit{An Introduction to Non-perturbative 
String Theory},  \texttt{arXiv:hep-th/9802051}.

\bibitem{prt} G. Papadopoulos, J.G. Russo and A.A. Tseytlin, \textit{Solvable 
model of strings in a time dependent plane wave background}, 
Class. Quant. Grav. 20 (2003) 969, \texttt{arXiv:hep-th/0211289}.

\bibitem{mmhom} M. Blau, M. O'Loughlin, \textit{Homogeneous Plane Waves},
Nucl. Phys. B654 (2003) 135-176, \texttt{arXiv:hep-th/0212135}.

\bibitem{bbop1} M. Blau, M. Borunda, M. O'Loughlin, G. Papadopoulos,
\textit{Penrose Limits and Spacetime Singularities}, Class. Quant. Grav.
21 (2004) L43--L49, \texttt{arXiv:hep-th/0312029}.  

\bibitem{bbop2} M. Blau, M. Borunda, M. O'Loughlin, G. Papadopoulos,
\textit{The universality of Penrose limits near space-time singularities},
JHEP 0407 (2004) 068, \texttt{arXiv:hep-th/0403252} 

\bibitem{gpsi}G. Papadopoulos, \textit{Power-law singularities in string 
theory and M-theory}, Class. Quant. Grav. 21 (2004) 5097-5120, 
\texttt{arXiv:hep-th/0404172}.

\bibitem{SI} P. Szekeres and V. Iyer, \textit{Spherically symmetric 
singularities and strong cosmic censorship}, Phys. Rev. D47 (1993) 4362.

\bibitem{bfw2} M. Blau, D. Frank, S. Weiss,
\textit{Fermi Coordinates and Penrose
Limits}, Class. Quantum Grav. 23 (2006) 3993-4010,
\texttt{arXiv:hep-th/0603109}.

\bibitem{mbsw} M. Blau, S. Weiss, \textit{Penrose Limits vs String
Expansions}, Class. Quantum Grav. 25 (2008) 125014,
\texttt{arXiv:0710.3480v1 [hep-th]} 

\bibitem{blrcbc} M. Blau and M. O'Loughlin, \textit{Multiple M2-Branes and
Plane Waves}, \texttt{arXiv:0806.3253 [hep-th]}.

\bibitem{bl} J. Bagger and N. Lambert, \textit{Modeling Multiple M2's}, 
Phys. Rev. D75 (2007) 045020,
\texttt{arXiv:hep-th/0611108};  J. Bagger and N. Lambert, 
\textit{Gauge symmetry and supersymmetry of multiple M2-branes}, 
Phys. Rev. D77 (2008) 065008,
\texttt{arXiv:0711.0955 [hep-th]};  J. Bagger and N. Lambert, 
\textit{Comments on multiple M2-branes}, JHEP 0802 (2008) 105, 
\texttt{arXiv:0712.3738 [hep-th]}.

\bibitem{gu} A. Gustavsson, \textit{Algebraic structures on 
parallel M2-branes}, \texttt{arXiv:0709.1260 [hep-th]}; A. Gustavsson,
\textit{Selfdual strings and loop space Nahm equations},
JHEP 0804 (2008) 083, \texttt{arXiv:0802.3456 [hep-th]}.

\bibitem{lor3} J. Gomis, G. Milanesi and J.G. Russo, \textit{Bagger-Lambert
Theory for General Lie Algebras}, \texttt{arXiv:0805.1012 [hep-th]};
S. Benvenuti, D. Rodriguez-Gomez, E. Tonni and H. Verlinde,
\textit{N=8 superconformal
gauge theories and M2 branes}, \texttt{arXiv:0805.1087 [hep-th]};
Pei-Ming Ho, Y. Imamura and Y. Matsuo, \textit{M2 to D2 revisited},
\texttt{arXiv:0805.1202 [hep-th]}.

\bibitem{dfsw} D. Frank and S. Weiss, to appear. 

\bibitem{dkps} M. Douglas, D. Kabat, P. Pouliot, S.Shenker, \textit{D-branes
and Short Distances in String Theory}, Nucl.Phys. B485 (1997) 85-127, 
\texttt{arXiv:hep-th/9608024v2}.

\bibitem{lifshytz} G. Lifshytz, \textit{DLCQ-M(atrix) Description of String 
Theory, and Supergravity}, 
Nucl. Phys. B534 (1998) 83, \texttt{arXiv:hep-th/9803191}.

\bibitem{penrose} R. Penrose, \textit{Any space-time has a plane wave as a
limit}, in \textit{Differential geometry and relativity}, Reidel, Dordrecht
(1976) pp.~271--275.

\bibitem{gueven} R. Gueven, \textit{Plane wave limits and T-duality},
Phys. Lett. B482 (2000) 255--263, \texttt{arXiv:hep-th/0005061}.

\bibitem{bfp} M. Blau, J. Figueroa-O'Farrill, G. Papadopoulos, 
\textit{Penrose limits, supergravity and brane dynamics}, 
Class. Quant. Grav. 19 (2002) 4753, \texttt{arXiv:hep-th/0202111}.

\bibitem{shomer} A. Shomer, \textit{Penrose limit and DLCQ of 
string theory}, Phys. Rev. D68 (2003) 086002, \texttt{arXiv:hep-th/0303055}.

\bibitem{sheikh} M.M. Sheikh-Jabbari, \textit{Tiny Graviton Matrix 
Theory: DLCQ of IIB Plane-Wave String Theory, A Conjecture},
JHEP 0409 (2004) 017, \texttt{arXiv:hep-th/0406214}.

\bibitem{torabian} M. Torabian, \textit{Matrix Theory for the DLCQ of 
Type IIB String Theory on the AdS/Plane-wave}, Phys. Rev. D76 (2007) 026006,
\texttt{arXiv:hep-th/0701046}.

\bibitem{marolfetal} D. Marolf, L. Martucci and P.J. Silva,
\textit{Actions and Fermionic symmetries for D-branes in bosonic backgrounds},
JHEP 0307 (2003) 019, \texttt{arXiv:hep-th/0306066}.

\bibitem{galloway} G.J. Galloway, \textit{Cosmological Spacetimes 
with $\Lambda > 0$}, \texttt{arXiv:gr-qc/0407100}.

\bibitem{brett} B. McInnes, \textit{Inaccessible Singularities in 
Toral Cosmology}, 
Class. Quant. Grav. 24 (2007) 1605, \texttt{arXiv:gr-qc/0611101}.

\bibitem{cht} B. Craps, T. Hertog, N. Turok, \textit{Quantum Resolution of
Cosmological Singularities using AdS/CFT}, \texttt{arXiv:0712.4180v2
[hep-th]}.

\bibitem{addnnt} A. Awad, S. Das, S. Nampuri, K. Narayan, S. Trivedi, \textit{Gauge Theories with Time Dependent
Couplings and their Cosmological Duals}, \texttt{arXiv:0807.1517v2 [hep-th]}.

\end{small}
\end{thebibliography}
\end{document}